\documentclass[12pt]{article}
\usepackage{amsmath,amssymb,amsthm,amscd}
\usepackage{epsfig}
\newcommand{\CN}{{\cal N}}
\newcommand{\CC}{{\cal C}}

\newcommand{\CO}{{\cal O}}

\newcommand{\CP}{{\cal P}}

\newcommand{\CR}{{\cal R}}

\def\IZ{{\mathbb Z}}
\def\IR{{\mathbb R}}
\def\IC{{\mathbb C}}
\def\IP{{\mathbb P}}

\def\IS{{\mathbb S}}
\newcommand{\tr}{{\rm Tr}}
\newcommand{\re}{{\rm e}}
\newcommand{\ri}{{\rm i}}

\newcommand{\e}{{\rm e}}
\newcommand{\dd}{{\rm d}}

\newcommand{\bea}{\begin{eqnarray}}
\newcommand{\eea}{\end{eqnarray}}
\newcommand{\beq}{\begin{equation}}
\newcommand{\eeq}{\end{equation}}
\newcommand{\be}{\begin{equation}}
\newcommand{\ee}{\end{equation}}
\newcommand{\ba}{\begin{aligned}}
\newcommand{\ea}{\end{aligned}}
\newcommand{\ben}{\begin{eqnarray}\displaystyle}
\newcommand{\een}{\end{eqnarray}}

\newcommand{\sectiono}[1]{\section{#1}\setcounter{equation}{0}}

\newdimen\tableauside\tableauside=1.0ex
\newdimen\tableaurule\tableaurule=0.4pt
\newdimen\tableaustep
\def\phantomhrule#1{\hbox{\vbox to0pt{\hrule height\tableaurule width#1\vss}}}
\def\phantomvrule#1{\vbox{\hbox to0pt{\vrule width\tableaurule height#1\hss}}}
\def\sqr{\vbox{%
  \phantomhrule\tableaustep
  \hbox{\phantomvrule\tableaustep\kern\tableaustep\phantomvrule\tableaustep}%
  \hbox{\vbox{\phantomhrule\tableauside}\kern-\tableaurule}}}
\def\squares#1{\hbox{\count0=#1\noindent\loop\sqr
  \advance\count0 by-1 \ifnum\count0>0\repeat}}
\def\tableau#1{\vcenter{\offinterlineskip
  \tableaustep=\tableauside\advance\tableaustep by-\tableaurule
  \kern\normallineskip\hbox
    {\kern\normallineskip\vbox
      {\gettableau#1 0 }%
     \kern\normallineskip\kern\tableaurule}%
  \kern\normallineskip\kern\tableaurule}}
\def\gettableau#1{\ifnum#1=0\let\next=\null\else
\squares{#1}\let\next=\gettableau\fi\next}

\newcommand{\figref}[1]{Fig.~\protect\ref{#1}}
\begin{document}
\begin{titlepage}

{}~
\hfill\vbox{
%\hbox{Parma--}
%\hbox{Florence--}
\hbox{CERN-PH-TH/2006-112}
}\break

\vskip .6cm

\centerline{\Large \bf
Phase transitions, double--scaling limit,}

\vspace*{1.0ex}
\centerline{\Large \bf and topological strings}

\medskip

\vspace*{4.0ex}

\centerline{\rm
Nicola Caporaso$^{a}$, Luca Griguolo$^{b},$ Marcos Mari\~no$^{c}$\footnote{Also at Departamento de
Matem\'atica, IST, Lisboa, Portugal.}, Sara Pasquetti$^{b}$ and Domenico Seminara$^{d}$ }

\vspace*{4.0ex}

\centerline{ \rm $^a$Center for Theoretical Physics, MIT, Cambridge, MA 02139 USA}
\centerline{\tt caporaso@fi.infn.it}

\vspace*{1.8ex}

\centerline{ \rm $^b$ Dipartimento di  Fisica, Universit\`a  di Parma,
INFN Gruppo Collegato di Parma,}
\centerline{\rm Parco Area delle Scienze 7/A, 43100 Parma, Italy}
\centerline{\tt  griguolo@fis.unipr.it, pasquetti@fis.unipr.it}

\vspace*{1.8ex}

\centerline{ \rm $^c$Department of Physics, Theory Division, CERN, Geneva 23, CH-1211 Switzerland}
\centerline{\tt
marcos@mail.cern.ch}
\vspace*{1.8ex}

\centerline{ \rm $^d$ Dipartimento di Fisica, Polo Scientifico Universit\`a di
Firenze,}
\centerline{\rm INFN Sezione di Firenze
Via  G. Sansone 1, 50019 Sesto Fiorentino, Italy}
\centerline{\tt seminara@fi.infn.it }

\vspace*{5ex}

\centerline{\bf Abstract}
\medskip
Topological strings on Calabi--Yau manifolds are known to undergo phase transitions at small distances. We
study this issue in the case of perturbative topological strings on local Calabi--Yau threefolds given
by a bundle over a two-sphere. This theory can be regarded as a q--deformation of Hurwitz theory, and it has a conjectural
nonperturbative description in terms of q--deformed 2d Yang--Mills theory. We solve the planar model and find a
phase transition at small radius in the universality class
of 2d gravity. We give strong evidence that there is a double--scaled theory at the critical point whose all genus free energy
is governed by the Painlev\'e I equation. We compare the critical behavior of the perturbative theory to the critical behavior of its
nonperturbative description, which belongs to the universality class of 2d supergravity, and we comment on possible 
implications for nonperturbative 2d gravity. We also
give evidence for a new open/closed duality relating these Calabi--Yau backgrounds to open strings with framing.

\end{titlepage}

\vfill
\eject

\tableofcontents

\sectiono{Introduction}

It has been recognized for a long time that string theory provides in a natural way a
deformation of classical geometry. The deformation parameter is given by
\be
\label{defpar}
{\ell_s\over R}
\ee
where $\ell_s$ is the lenght of the string and $R$ is the characteristic size of the target space.
When this parameter is very small, the target geometry can be regarded as a classical
background corrected by stringy effects. However, when this parameter is big, classical geometric intuition breaks down and
one is forced to use some notion of stringy or quantum geometry.

These phenomena have been studied in detail in the context of type A topological string theory on Calabi--Yau manifolds, where the characteristic
size $R$ of the target space is set by the K\"ahler moduli (see \cite{greene, aspinwall} for reviews and a list of references).
When these sizes are small, so that the deformation parameter (\ref{defpar}) is big,
the breakdown of classical geometry can be made precise quantitatively by looking at the behavior of topological string amplitudes. Typically, the large radius
expansion of the amplitudes becomes divergent at a critical value of the K\"ahler moduli, and this divergence signals the onset of the stringy regime.
We will refer to this change of regime as a phase transition. In known cases (like the quintic), this transition exhibits a universal behavior which has been identified
with the $c=1$ string at the
self--dual radius \cite{bcov, ghv}.

In this paper we will analyze the critical behavior of type A topological string theory in the local Calabi--Yau manifold
\be
X_p=\CO(p-2) \oplus \CO(-p) \rightarrow \IP^1.
\ee
This theory has been studied from a mathematical point of view in
\cite{bp} and it is closely related to Hurwitz theory. From a
physical point of view, it was proposed in \cite{aosv} that the
model could be defined non--perturbatively by using a q--deformed
version of 2d Yang--Mills theory. Although perturbative topological
string theory on the background $X_p$ shares many known properties
with other local Calabi--Yau manifolds (like for example
Gopakumar--Vafa integrality), we will show in this paper that for
$p>2$ it exhibits a phase transition at small radius which is not in
the same universality class as the examples considered so far, i.e.
the $c=1$ string. Rather, the critical behavior is in the
universality class of pure two--dimensional gravity (i.e. the
$(2,3)$ model) for all $p>2$.

It turns out that, as in the case of matrix models, the critical behavior of higher genus amplitudes makes it possible to define a double--scaled theory at the
transition point. The double--scaling limit erases a lot of information about the original topological string theory, since it only keeps the leading singularity in the
genus $g$ free energies, but it is interesting for at least three reasons:

(1) It gives a compact characterization of the critical behavior at every order in string perturbation theory.

(2) It makes possible to study the free energy at all genus, and provides information about the nonperturbative structure of the theory.

(3) As argued in \cite{liu,ag} in the context of unitary matrix models
describing thermal AdS, the phase transition might be
an artifact of perturbation theory, and the double--scaling limit provides a way to smooth out the critical behavior.

We give convincing evidence that the double--scaled free energy is
described  for all $p>2$ by the Painlev\'e I equation which characterizes 2d gravity \cite{gmigdal,bk,ds}.
The resulting scaled free energy is known to be non--Borel summable as a genus expansion, and this confirms that
perturbative topological string theory on these backgrounds lacks fundamental non--perturbative input. Interestingly, the
non--perturbative definition of the model proposed in \cite{aosv} also exhibits critical behavior \cite{capo,abms, jm}, but this time
in the universality class of the Gross--Witten--Wadia unitary model \cite{gw,w} described by Painlev\'e II \cite{ps,gm}, which is also
the universality class of 2d supergravity \cite{kms}. The Painlev\'e II equation
is known to have a well--defined unique solution with the appropriate boundary
conditions \cite{cdm}, and this can be regarded as further support for the proposal of \cite{aosv} as a non--perturbative definition of the theory. 

Conversely,
our result shows that 2d gravity and 2d supergravity can be respectively embdedded in a one--family of topological string theories and their nonperturbative holographic completions. The relation that emerges between 2d gravity and 2d supergravity can be better understood by 
using the results of \cite{djm, djmw}, and we speculate that the proposal of \cite{aosv}, together with our results on the critical behavior, 
give further support to the nonperturbative solution of 2d gravity presented in \cite{djm}.

Mathematically, perturbative string theory on $X_p$ can be regarded as a q--deformation of Hurwitz theory, and the Gromov--Witten invariants of $X_p$ promote
simple Hurwitz numbers to polynomials. This ingredient is very important in our study, and indeed the critical behavior we describe for $p>2$ is already
present in the Hurwitz case. This was noted at the planar level in \cite{ksw}, and it is implicit in the results of \cite{gjv} for higher genus Hurwitz theory. Our results are also
probably related to the description of topological gravity in terms of the asymptotics of Hurwitz numbers due to Okounkov and Pandharipande \cite{op}. In fact, we can reach
higher critical points by perturbing the model with higher Casimir operators. This perturbed model is an analog of the general one--matrix model, while the theory
of Hurwitz numbers should be
an analog of the Gaussian model with general D--brane insertions. Both are related by an open--closed duality in the spirit of \cite{adkmv,mmss}.

Double--scaling limits of topological string theory
have been studied before, mostly in the context of the Dijkgraaf--Vafa correspondence (see \cite{miramontes} for a recent example with references to the literature)\footnote{\cite{dg}
considers the double--scaling limit of Chern--Simons theory on the sphere, and recovers the $c=1$ behavior at the conifold.}.
In that case one studies type B topological string theory on Calabi--Yau backgrounds related to the deformed conifold. The topological string amplitudes are
given by matrix integrals, which can then be analyzed with the standard technology developed in the study of two--dimensional
gravity. In contrast, we look at type A topological strings in backgrounds related to the resolved conifold, where the amplitudes are given by sums over partitions, and the
connection to 2d gravity is more surprising (although it could have been suspected with some hindsight, due to the results of \cite{ksw,gjv,op}).

The organization of this paper is as follows. In section 2 we
review some results on phase transitions and phases of topological
string theory. In section 3 we introduce perturbative topological
string theory on $X_p$, write down its partition function, and
relate it to various close cousins, in particular, we show that
the model can be regarded in a precise sense as a q--deformation
of Hurwitz theory. In section 4 we solve the model at genus zero
(i.e. the planar theory) by using standard matrix model
techniques. In section 5 we extract the critical behavior of the
planar theory and we show that including higher Casimirs leads to
multicritical points a la Kazakov (albeit we do the detailed
analysis only in the undeformed case, i.e. in Hurwitz theory). In
section 6 we analyze the higher genus theory. To do this we make
an ansatz for $F_g(t)$ which generalizes the results of Goulden,
Jackson and Vakil for Hurwitz theory \cite{gjv,gj} to the deformed
case. We also define the double--scaled theory and give strong
evidence that it is described by 2d gravity. In section 7 we
compare the structure found in the perturbative setting with the
structure that one finds in the non--perturbative definition given
by q--deformed 2d YM, and we speculate on the possible consequences 
for 2d gravity. In section 8 we point out evidence for a new
open/closed duality generalizing the results of Aganagic and Vafa
in \cite{av}. In Appendix A we recall the very useful Lagrange
inversion formula and apply it to various examples of the paper.
In Appendix B we give some useful integrals for the saddle--point
analysis of section 4.

\sectiono{Phase transitions in topological string theory}

For simplicity, we will assume that the Calabi--Yau $X$ has a single K\"ahler parameter $t$, i.e. $h^{1,1}(X)=1$. This is usually taken as a complexified
parameter of the form
\be
t=r+\ri \theta,
\ee
but since we will be interested in questions regarding the convergence of the prepotential as a function of $\re^{-t}$, the $\theta$ parameter is not relevant and we will set it to zero.

\subsection{Phase transitions in the planar free energy}

As we mentioned
in the introduction, when $t$ is large (in the so--called large radius regime) the geometry probed by string theory can be regarded as a classical geometry together with
stringy corrections. This is well reflected in the structure of the prepotential $F_0 (t)$ or genus zero topological string amplitude, which in the large radius regime is of the
form
\be
\label{prep}
F_0(t) ={C\over 6}t^3 +\sum_{k=1}^{\infty} N_{0,k} \re^{-kt}.
\ee
In this equation, $C$ is the classical intersection number for the two-cycle whose size is measured by $t$. The infinite sum in the r.h.s. is given by worldsheet instanton
corrections, which are obtained by ``counting" (in an appropriate sense) holomorphic maps from $\IP^1$ to $X$. The instanton counting numbers $N_{0,k}$ are
genus zero Gromov--Witten invariants, and we have chosen units in which $\ell_s = {\sqrt{2\pi}}$.

The series of worldsheet instanton corrections, regarded as a power series in $\re^{-t}$, has in general a finite radius of convergence $t_c$
which can be obtained by looking at the asymptotic growth with $k$
of the numbers $N_{0,k}$. We will characterize this asymptotic growth by $t_c$ and by a critical exponent $\gamma$:
\be
\label{asymprep}
N_{0,k} \sim  k^{\gamma -3} \re^{k t_c}, \quad k \rightarrow \infty.
\ee
When this holds, the prepotential behaves near $t_c$ as
\be \label{critprep} F_0(t) \sim (\re^{-t_c} - \re^{-t})^{2-\gamma}.
\ee
It turns out that typical Gromov--Witten invariants of Calabi--Yau manifolds behave asymptotically as
\be
\label{candelaslike}
N_{0,k} \sim {\re^{kt_c} \over  k^3 \log^2\, k}, \quad k \rightarrow \infty.
\ee
This is of the form (\ref{asymprep}), with critical exponent
\be
\gamma=0
\ee
and subleading log corrections.
This behavior was first established in \cite{candelas} in the example of the quintic, and since then it
has been verified in other examples, like for example in local $\IP^2$, where the critical radius is given by
 \cite{msd,kz}
 \be
 t_c = {1\over \Gamma\Bigl( {1\over 3} \Bigr)   \Gamma\Bigl( {2 \over 3} \Bigr)}{\rm Re}\, G\Bigl({1\over 3}, {2\over 3}, 1;1\Bigr) \sim 2.90759
 \ee
 and $G$ is the Meijer function.

The subleading $\log$ in (\ref{candelaslike}) leads to log corrections near the critical point (also referred to as scaling violations)
of the form
\be
F_0(t) \sim  (\re^{-t_c} - \re^{-t})^2 \log (\re^{-t_c} - \re^{-t}).
\ee
This is the genus zero free energy of the $c=1$ string at the self--dual radius, once the scaling variable $\re^{-t_c} - \re^{-t}$ is identified with the cosmological constant \cite{bcov, ghv}.

The behavior of the prepotential gives a precise quantitative meaning to the distinction between classical and quantum geometry. We will refer to the divergence of the large radius
expansion at $t=t_c$ as a {\it phase transition} with a critical exponent $\gamma$ defined in (\ref{candelaslike}). The phase with
\be
t > t_c
\ee
where the expansion (\ref{prep}) is convergent, is called the large radius or Calabi--Yau phase, where classical geometry makes sense (albeit it is corrected by
worldsheet instantons). When $t\le t_c$, the nonlinear sigma model approach is not well defined, and classical geometric
intuition is misleading.

There are two possible approaches to this problem. In \cite{msd,mm}, Aspinwall, Greene and Morrison proposed to study this small area phase by using mirror symmetry.
One finds that the point $t=t_c$ separating the two phases corresponds to the discriminant locus of the mirror moduli space (the conifold point). For
$t<t_c$, mirror symmetry makes possible to compute the prepotential by analytic continuation. The second approach is due to Witten, who proposed in \cite{phases}
the linear $\sigma$ model description of $\CN=2$ sigma models on Calabi--Yau manifolds largely as a way to understand this phase transition. In this formalism, one
recovers the mirror symmetric description, and
in addition the stringy or non--geometric phase
can be described in a precise way. It is typically a Landau--Ginzburg orbifold perturbed by twist operators.

\subsection{Higher genus}

In order to describe the phase structure of the model we have relied on the behavior of the prepotential, i.e. the planar free energy.
It is natural to ask what happens when higher genus topological string amplitudes are taken into account.
These amplitudes, which we denote by
$F_g(t)$, can be expressed in the large radius limit
in terms of genus $g$ Gromov--Witten invariants $N_{g,k}$:
 \be
 F_g(t) =\sum_{k=1}^{\infty} N_{g,k}\re^{-kt}.
 \ee
 Here we have omitted contributions from degree zero maps as well as classical pieces for $g=1$, which will not be important
 in the discussion. It turns out that the Gromov--Witten invariants have the asymptotic behavior \cite{bcov}
 \be
 \label{gwgenusg}
N_{g,k}\sim k^{(\gamma -2)(1-g) -1} \re^{kt_c}, \quad k \rightarrow \infty,
 \ee
 where $t_c$ is the critical radius obtained at genus zero and it is common
 to all $g$, and $\gamma$ is the critical exponent that appears in (\ref{asymprep}). This is equivalent to the following behavior near the critical point
\be
\label{fgcrit}
\ba
F_1(t) &\sim c_1 \log\, (\re^{-t_c} -\re^{-t}),\\
 F_g(t) & \sim c_g (\re^{-t_c} -\re^{-t})^{(1-g)(2-\gamma)}, \quad g\ge 2.
 \ea
\ee
  In conventional topological string theory, as we have mentioned, $\gamma=0$, but the more general form we have written above
 will be useful later.

 We then see that the phase transition at $t=t_c$ is common for all $F_g(t)$, and the critical exponent
 changes with the genus in the way prescribed by (\ref{gwgenusg}). This sort of coherent behavior in the genus expansion is not obvious, but
 seems to characterize a wide variety of systems that admit a genus expansion (like for example matrix models, see \cite{dfgz} for a review).
 When this is the case, one can define a {\it double--scaling limit} \cite{gmigdal,bk,ds} as follows. Let us consider the total free energy $F$ as a perturbative
expansion in powers of the string coupling constant $g_s$:
\be
\label{fallgenus}
F(g_s, t) =\sum_{g=0}^{\infty} F_g(t) g_s^{2g-2}.
\ee
 We define the double--scaled string coupling as
 \be
 \kappa= a  g_s (\re^{-t_c} -\re^{-t})^{\gamma/2-1},
 \ee
 where $a$ is an appropriate constant. We can then consider the limit
 \be
 t\rightarrow t_c, \qquad g_s \rightarrow 0, \qquad \kappa \,\, {\rm fixed}.
 \ee
 In this limit, only the most singular part of $F_g(t)$ survives at each genus, and the total free energy becomes the
 {\it double-scaled free energy}
 \be
 F_{\rm ds}(\kappa)=f_0 \kappa^{-2} + f_1 \log \, \kappa + \sum_{g\ge 2} f_g \kappa^{2g-2},
 \ee
 where $f_g=a^{2-2g} c_g$. It is also customary to express the double--scaled free energy in terms of the
 scaling variable $z= \kappa^{2/(\gamma -2)}$.

 It turns out that, in some cases, one can determine the coefficients $f_g$ in closed form. In the double--scaling limit of
 matrix models, they are governed by a differential equation of the Painlev\'e type \cite{dfgz}. In the case of topological string theory on
 Calabi--Yau manifolds, it was conjectured in \cite{ghv} that, in terms of a natural coordinate
 \be
 \mu \sim \re^{-t_c} -\re^{-t}
 \ee
 which in the mirror model measures the distance to the conifold point $\mu=0$, the double-scaled free energy is universal and reads
 \be
 \label{cone}
 F_{ds}(\mu) ={1\over 2} \mu^2 \log \, \mu -{1\over 12} \log \, \mu+   \sum_{g=2}^{\infty} {B_{2g} \over 2g (2g-2) } \mu^{2-2g}.
 \ee
 This is exactly the all genus free energy of the $c=1$ string at the self--dual radius (for a review, see \cite{klebanov}). This
 behavior has been checked in many examples \cite{kz,kkv}.

\sectiono{Topological string theory on bundles over $\IP^1$}

\subsection{The model and its partition function}

In this paper we study perturbative topological string theory on a Calabi--Yau threefold given by the total space of a
rank two holomorphic complex
bundle over $\IP^1$. It is a well--known result that these bundles split into a direct sum of line bundles, therefore these spaces are of the form
\be
\CO(p-2) \oplus \CO(-p) \rightarrow \IP^1,
\ee
where $p \in \IZ$. There is an obvious symmetry
\be
\label{psym}
p \rightarrow -p+2
\ee
and we can restrict to $p>0$.

The topological string partition function on the space $X_p$ is the exponential of the total free energy (\ref{fallgenus})
\be
Z_{X_p}=\exp \, F_{X_p}(g_s,t)
\ee
where
\be
\label{totalfgen}
F_{X_p}(g_s,t) =\sum_{g=0}^{\infty} g_s^{2g-2} F^{X_p}_g(t)=\sum_{g=0}^{\infty}\sum_{d=1}^{\infty} N_{g,d}(p) \re^{-dt} g_s^{2g-2},
\ee
and $N_{d,g}(p)$ are the Gromov--Witten invariants. This partition function can be computed by using
the topological vertex \cite{akmv}, and the same result is obtained from the local Gromov--Witten theory of curves of
\cite{bp} (see section 2 of \cite{aosv} for a useful summary).
In order to write down the result,
we recall some necessary ingredients. First of all, we define the q--number $[n]$ as
\be\label{qnumb} [n]=q^{n/2} -q^{-{n/2}}, \qquad q=\re^{-g_s}. \ee
A representation $R$ of $U(\infty)$ can be represented by a Young tableau, labeled by the lengths of its rows $\{l_i\}$. The
quantity
\be
\ell(R) =\sum_i l_i
\ee
is the total number of boxes of the tableau. Notice that a tableau $R$ with $\ell(R)$ boxes can be regarded as a representation of the
symmetric group $S_{\ell(R)}$ of $\ell(R)$ elements. Another important quantity associated to a tableau is
\be
\kappa_R=\sum_i l_i(l_i-2i+1),
\ee
which is closely related to the second Casimir of $R$ regarded as a $U(N)$ representation. Finally, we introduce the quantity
\be
W_R =q^{-\kappa_R/4} \prod_{\tableau{1} \in R} {1\over [{\rm hook}(\tableau{1})]}.
\ee
This is in fact a specialization of the topological vertex. The product in the r.h.s. is over all the boxes in the tableau, and hook$(\tableau{1})$ denotes
the hook length of a given box in the tableau.

The topological string partition function on $X_p$ is given by
\be
\label{totalz}
Z_{X_p}=\sum_R W_R W_{R^t} q^{(p-1) \kappa_R/2}(-1)^{\ell(R)p} \re^{-\ell(R) t}
\ee
where $R^t$ denotes the transpose tableau (i.e. the tableau where we exchange rows and columns). The above expression has to be understood, as
in \cite{akmv}, as a power series in $\re^{-t}$. At every degree it gives the all--genus answer. One can easily compute the first few terms in the expansion:
\be
\label{foex}
\ba
F^{X_p}_0(t) &= (-1)^p \re^{-t} +{1\over 8}(2 \,  p^2 -4\, p +1) \re^{-2t} +{(-1)^p\over 54} (1-6\, p + 3\, p^2) (2-6\, p + 3\, p^2)\re^{-3t} \\&
+\CO(\re^{-4t}),\\
F^{X_p}_1(t)&=-{(-1)^p \over 12}  \re^{-t} +{1\over 48} (p^4 -4 \, p^3 +p^2 +6\, p -2) \re^{-2t} \\& +{(-1)^p \over 72}
 (-2 + 14\,p - 19\,p^2 - 20\,p^3 + 45\,p^4 - 24\,p^5 + 4\,p^6)\re^{-3t} + \CO(\re^{-4t}),
 \ea
 \ee
 and so on. The Gromov--Witten invariants $N_{g,d}(p)$ are polynomials in $p$ of degree $2d-2+2g$ with
 rational coefficients:
 \be
 \label{gwpol}
 N_{g,d}(p) =\sum_{i=0}^{2d-2+2g} N^i_{g,d} \, p^i.
 \ee
 As we will see, the coefficient of the highest power in (\ref{gwpol}) has a simple geometric interpretation in terms
 of classical Hurwitz theory. Note that, due to the definition we are using of $g_s$, the Gromov--Witten invariants
 $N_{d,g}$ differ in a sign $(-1)^{g-1}$ from the standard ones. This is of no consequence for most of our
 discussion, but the extra sign should be taken into account when extracting integer Gopakumar--Vafa invariants $n_{g,d}(p)$ \cite{gv} from the above
 expressions. 

The partition function (\ref{totalz}) on $X_p$ is calculated, as explained in \cite{bp}, by considering the equivariant Gromov--Witten theory on $X_p$
 with respect to scalings of the line bundles over $\IP^1$, and by using the antidiagonal action. This is, therefore, an equivariant partition function, and leads to a
 topological string theory which is different from the standard ones. For $p=1$ this theory specializes to (standard) topological string theory on the resolved conifold.
 For $p=2$, the geometry is that of $\IC \times A_1$. The standard topological string theory has a trivial partition function on this space, while the
 equivariant theory considered here has $Z_{X_2}=Z_{X_1}^{-1}$. See \cite{fj} for more details on the equivariant theory from the point of view of mirror symmetry.

 Another way to understand the geometric content of this non--standard theory is to consider the equivariant version of topological strings presented in
section 7.6 of \cite{amv}. Let us look at equivariant topological string theory on local $\IP^2$. This model contains, on top of the K\"ahler parameter $t$,
three equivariant parameters $x_i$, $i=1, 2,3$, related to twisted masses in the sigma model. The partition function depends now on the $x_i$,
\be
\label{equivptwo}
Z_{\IP^2}(x_1, x_2, x_3),
\ee
and can be computed again by using the topological vertex (explicit formulae are given in \cite{amv}). It is completely symmetric under permutations of the $x_i$ and,
if we set $x_i=1$ we recover the standard local $\IP^2$ partition function. One can deduce from the explicit formulae in \cite{amv} that
\be
Z_{\IP^2}(1, 0, 0) =Z_{X_3}.
\ee
This is easy to understand, since the hyperplane class of $\IP^2$ in the total geometry is indeed a $\IP^1$ with normal bundle $\CO(-3) \oplus \CO(1)$. By taking $x_2=x_3=0$
we are considering the local geometry of this curve inside $\CO(-3) \rightarrow \IP^2$. For $p>3$ we can regard $Z_{X_p}$ as a specialization of equivariant topological string theory on more
complicated toric backgrounds. 

Finally, we point out that (\ref{totalz}) can be interpreted in terms of representations of Hecke algebras by using the results of \cite{deharo}.

 \subsection{Relation to Hurwitz theory}

 Before analyzing this partition sum in detail, let us relate it to other generating functionals. The quantity $W_R$ turns out to be a q--deformation
 of the dimension $d_R$ of the representation $R$ of $S_{\ell(R)}$. As $g_s \rightarrow 0$, one has that
 \be
W_R \rightarrow g_s^{-\ell(R)} {d_R \over |\ell(R)|!}.
\ee
This suggests taking the following limit,
\be
\label{double}
g_s \rightarrow 0, \quad t\rightarrow \infty, \quad p \rightarrow \infty,
\ee
in such a way that
\be
\label{scaling}
p g_s =\tau_2/N, \quad (-1)^{p}  \re^{-t}=(g_s N)^2 \re^{-\tau_1},
\ee
and $\tau_1$, $\tau_2$ and $N$ are new parameters that are kept fixed. It will be convenient in what follows to introduce a 't Hooft parameter as
\be \label{hooft} T=g_s N. \ee
In the limit (\ref{double})--(\ref{scaling}) one has $T\rightarrow 0$. The partition function becomes:
\be\label{zksw} Z_{X_p} \rightarrow Z_{\rm Hurwitz}=\sum_R\biggl(
{d_R \over |\ell(R)|!}\biggr)^2 N^{2\ell(R)} \re^{-\tau_2
\kappa_R/2N} \re^{-\tau_1\ell(R)} . \ee
This is the generating functional of simple Hurwitz numbers of $\IP^1$ at all genus and degrees.
In order to see this, it is useful to recall some basic ingredients of Hurwitz theory.

Hurwitz theory studies branched covers of Riemann surfaces. The structure of the
covering will be labeled by a vector $\vec k$ of nonnegative entries, and we define
\be
\ell(\vec k)=\sum_j j k_j, \quad |\vec k| =\sum_j k_j, \quad z_{\vec k} =\prod_j j^{k_j} k_j!.
\ee
If a cover has degree $d$, one necessarily has
\be
\ell(\vec k)=d.
\ee
Given $m$ branched points on $\IP^1$,
their branching structure can be specified by $m$ vectors
\be
\vec k^{(1)}, \cdots, \vec k^{(m)}.
\ee
The Riemann--Hurwitz formula relates the entries of these vectors, the degree of the cover and the
genus $g$ of the covering surface as follows:
\be
2g-2 +2d = \sum_i (\ell(\vec k ^{(i)}) - |\vec k^{(i)}| ).
\ee
The number of Hurwitz covers at degree $d$ is given by the classical formula
\be
H^{\IP^1}_d (\vec k_1, \cdots ,\vec k_m) =\sum_{\ell(R)=d} \biggl( {d_R \over \ell(R)!}\biggr)^2 \prod_{i=1}^m  f_R(\vec k^{(i)}).
\ee
In this equation,
\be
f_R(\vec k)= {\ell(R)!\over d_R} {\chi_R(C(\vec k)) \over z_{\vec k}},
\ee
where $C(\vec k)$ is the conjugacy class associated to the vector $\vec k$ in the symmetric group of $\ell(\vec k)$ elements, and
$\chi_R(C(\vec k))$ is its character.

A branched point whose branching vector is of the form
\be
\vec k_i=(d-2,1,\cdots)\equiv (2)
\ee
is called a simple branched point. A particularly important example of Hurwitz theory occurs when we have one non--simple branched point specified
by a vector $\vec k$, and $m(g, \vec k)$ simple branched points. By Riemann--Hurwitz, the number of simple branched points is given by
\be
m(g, \vec k)=2g-2+|\vec k| +\ell(\vec k),
\ee
where we used that $d=\ell(\vec k)$. Since
\be
 f_R(d-2,1,0,\cdots)={1\over 2}\kappa_R,
 \ee
The Hurwitz number for this configuration is given by
\be
\label{nshurwitz}
H_{g,d}^{\IP^1} (\vec k) =\sum_{\ell(R)=d} \biggl( {d_R \over \ell(R)!}\biggr)^2 f_R(\vec k) (\kappa_R/2)^{m(g,\vec k)}.
\ee

The case where there are only $m$ simple branched points is a particular case of the above. It is obtained when the branching
structure is trivial,
\be
\vec k =(d,0, \cdots) \equiv 1^d,
\ee
and
\be
\label{rh}
m=2g-2+2d.
\ee
The resulting Hurwitz number is called a {\it simple Hurwitz number}, and it is given at genus $g$ and degree $d$ by
\be
H_{g,d}^{\IP^1} (1^d) =\sum_{\ell(R)=d} \biggl( {d_R \over \ell(R)!}\biggr)^2  (\kappa_R/2)^{2g-2+ 2d},
\ee
where the sum is over representations $R$ with fixed number of boxes equal to the degree $d$.

We can now rewrite (\ref{zksw}) as
\be
\ba
Z_{\rm Hurwitz}&=\sum_{d,m}  N^{2d-m} \re^{-\tau_1 d} \sum_{\ell(R)=d}\biggl( {d_R \over \ell(R)!}\biggr)^2 { (- \tau_2)^m \over m!} (\kappa_R/2)^m \\
&=\sum_{g\ge 0} N^{2-2g} \sum_{d\ge 0} \re^{-\tau_1 d}  H_{g,d}^{\IP^1} (1^d) { \tau_2^{2g-2+2d} \over (2g-2+2d)!},
\ea
\ee
where in the second line we have traded the sum over $m$ by a sum over $g$, and we used (\ref{rh}).
The model described by (\ref{zksw}) has been studied in detail due to its connection to Hurwitz theory. From the physical
 point of view, it was analyzed in \cite{ksw,t, ct}, and
in the mathematical literature it has been studied for example in
\cite{gjv,gj}.

The free energy of $Z_{\rm Hurwitz}$ describes {\it connected}, simple Hurwitz numbers $H_{g,d}^{\IP^1} (1^d)^{\bullet}$:
\be
\label{freehurwitz}
F_{\rm Hurwitz}=\log Z_{\rm Hurwitz}= \sum_{g\ge 0} N^{2-2g} \sum_{d\ge 0} \re^{-\tau_1 d}  H_{g,d}^{\IP^1} (1^d)^{\bullet} { \tau_2^{2g-2+2d} \over (2g-2+2d)!}
\ee
If we compare this to the total free energy $F_{X_p}$ written in (\ref{totalfgen}) in terms of Gromov--Witten invariants,
and take the limit (\ref{double})--(\ref{scaling}), we find
\be
\label{hurlimit}
\lim_{p\rightarrow \infty} p^{2-2g-2d} (-1)^pN_{g,d}(p)= { H^{\IP^1}_{g,d}(1^d)^{\bullet}\over (2g-2+2d)!}.
\ee
 The l.h.s. is precisely the coefficient of the highest power in $p$ of (\ref{gwpol})\footnote{This result, for the genus zero case, was already derived in
\cite{toptwo}}. We can therefore interpret the Gromov--Witten invariants of this model as
 q--deformed connected, simple Hurwitz numbers, since they promote $H^{\IP^1}_{g,d}(1^d)^{\bullet}$ to polynomials of degree $2g-2+2d$ (which is equal to the number of
simple branch points).

We also point out that a closely related model to the partition function $Z_{X_p}$ and its Hurwitz limit
is $U(1)$, four--dimensional $\CN=2$ supersymmetric gauge theory with Casimir operators, considered in \cite{lmn} from the point of 
view of Nekrasov's instanton counting \cite{nekrasov}. When only the first
and the second Casimir operators are turned on, this
model is equivalent to the Hurwitz model (\ref{zksw}), and via the Hurwitz/Gromov--Witten correspondence of \cite{optwo}, it is related to topological string theory on $\IP^1$.
The q--deformed model $Z_{X_p}$ we are considering here is in turn related to the five--dimensional or K--theoretic version of the gauge theory
(see for example \cite{maeda}).

 \subsection{Relation to open Gromov--Witten invariants}

There is a generating functional in {\it open} Gromov--Witten theory which is also closely related to Hurwitz numbers.
Consider a Lagrangian submanifold with the topology $\IC \times \IS^1$ in $\IC^3$, and with framing $f$. The generating functional of
open Gromov--Witten invariants can be computed with the Chern--Simons invariants of the unknot \cite{ov} or with the topological vertex \cite{akmv,m}, and reads:
\be
\label{open}
Z(V)=\sum_R W_R q^{f \kappa_R/2} (-1)^{\ell(R)f} \tr_R\, V,
\ee
where $V$ is a $U(\infty)$ matrix source of open string moduli that takes into account the open string sectors associated to
different winding numbers. Let us now rescale $V\rightarrow \re^{-t} V$
and take the limit (appropriate for the open sector)
\be
g_s \rightarrow 0, \quad t\rightarrow \infty, \quad f\rightarrow \infty,
\ee
keeping fixed
\be
\label{scalingtwo}
f g_s =\tau_2/N, \quad (-1)^{f}\re^{-t} =g_s N \re^{-\tau_1}.
\ee
The generating functional (\ref{open}) becomes
\be
\label{genhgf}
\ba
Z_{\rm Hurwitz} (V)&= \sum_R N^{\ell(R)}  \biggl( {d_R \over \ell(R)!}\biggr) \re^{-\tau_2 \kappa_R/2N-\tau_1 \ell(R)} \tr_R V \\
&=1+ \sum_{\vec k} \sum_{g\ge 0} N^{2-2g-|\vec k|} \re^{-\tau_1 \ell(\vec k)}  H_{g,\ell(\vec k)}^{\IP^1} (\vec k) { (- \tau_2)^{2g-2+|\vec k| +\ell(\vec k)} \over (2g-2+ |\vec k| +\ell(\vec k))!} \Upsilon_{\vec k}(V),
\ea
\ee
where
\be
\Upsilon_{\vec k}(V) =\prod_j \bigl( \tr\, V^j \bigr)^{k_j}.
\ee
(\ref{genhgf}) is a generating functional for the more general Hurwitz numbers (\ref{nshurwitz}). If we now consider the free energy associated to the
open Gromov--Witten functional, and we write them in terms of open Gromov--Witten invariants $F_{g, \vec k}(f)$,
\be
F(V) =\sum_{\vec k} \sum_{g=0}^{\infty} F_{g,\vec k} (f) g_s^{2g-2+|\vec k|} \Upsilon_{\vec k}(V),
\ee
we find that
\be
\label{openrel}
\lim_{f\rightarrow \infty} f^{2-2g-|\vec k| - \ell(\vec k)}  (-1)^f  F_{g,\vec k} (f) ={H_{g,\ell(\vec k)}^{\IP^1} (\vec k)^{\bullet}\over (2g-2+ |\vec k| +\ell(\vec k))!}.
\ee
This expresses the more general Hurwitz numbers (\ref{nshurwitz}) as limits of open Gromov--Witten invariants. We point out that the open Gromov--Witten invariants
$F_{g,\vec k}(f)$ can be expressed in terms of triple Hodge integrals \cite{kl,mv}, while the Hurwitz numbers (\ref{nshurwitz}) can be written in terms of simple Hodge integrals
\cite{elsv}, and the relation (\ref{openrel}) has been noted in \cite{ophodge, llz} by considering their integral expression.

We now relate the closed Gromov--Witten invariants $N_{g,d}(p)$ in the background $X_p$ to the open Gromov--Witten invariants $F_{g, \vec k}(f)$.
It is clear that (\ref{open}) reduces to (\ref{totalz}) if we set $f=p-1$ and if we give a value to the source $V$ in such a way that
\be
\label{vcondensation}
\tr_R \, V = (-1)^{\ell(R)}\re^{-\ell(R) t} W_{R^t}.
\ee
There is indeed a choice of the (infinite--dimensional) matrix $V$ which produces this, with eigenvalues $x_i=q^{i-1/2}\re^{-t}$.
With this choice, one has that
\be
\label{vback}
\Upsilon_{\vec k}(V) =(-1)^{ |\vec k|} \re^{-\ell(\vec k) t} \prod_j {1\over (q^{j\over 2} -q^{-{j\over 2}})^{k_j}}
\ee
If we expand this in powers of $g_s$,
\be
\Upsilon_{\vec k}(V)=\re^{-\ell(\vec k) t}\sum_{g=0}^{\infty} g_s^{-|\vec k| + 2g} c_{g, \vec k},
\ee
where $c_{g,\vec k}$ are rational numbers, we find
\be
N_{g,d}(p) =\sum_{\ell(\vec k)=d} \sum_{g'=0}^{g} F_{g',\vec k}(p) c_{g-g',\vec k} ,
\ee
which can be used to express the q--deformed Hurwitz number $N_{g,d}(p)$ in terms of triple Hodge integrals.

There is a more elegant way to relate the closed and the open invariants, by using the underlying integer invariants. It was shown in
section 5 of \cite{amv} that, when closed string functionals are obtained from open string functionals by ``condensing" the source term as in
(\ref{vcondensation}), the Gopakumar--Vafa invariants $n_{g,d}$ of the closed geometry can be written in terms of the open BPS invariants introduced in
\cite{ov,lmv}. In the case of framed branes in $\IC^3$, the integer invariants $N_{R,g}(f)$ depend only on the genus $g$
and a representation $R$ of $U(\infty)$. It is easy to obtain from the results in \cite{amv} that
\be
\label{openclosed}
n_{g,d}(p) = N_{\underbrace{\tableau{2} \cdots \tableau{2}}_{d \, {\rm boxes}}, g}(p-1).
\ee
The tableau in the r.h.s. corresponds to the trivial representation
of $S_d$, and in deriving this relation we have used the conventions
for open BPS invariants in \cite{mv,lmv}.

\sectiono{The planar limit}

The explicit expression (\ref{totalz}) gives the total partition function but it is not useful in extracting the functions
$F_g^{X_p}(t)$ in a compact way. If we are interested for example in the asymptotic behavior of the Gromov--Witten invariants for large degree,
we would like to have closed expressions for the free energies at fixed genus but at all degrees. These are the kind of expressions that
can be obtained by using mirror symmetry. Unfortunately, there is no complete mirror description of the $X_p$ backgrounds that can be
used efficiently to compute $F_g^{X_p}(t)$. Some results concerning this description have been recently obtained in \cite{fj}, but most of the general
ingredients (like Picard--Fuchs equations) are still lacking.

In these circumstances, we have to extract the $F_g^{X_p}(t)$ directly from the sum over partitions (\ref{totalz}). In this section we solve the planar
limit, i.e. we find the genus zero contribution or prepotential, by doing a saddle--point analysis of (\ref{totalz}). Some aspects of our analysis were
 worked out in
\cite{jm}.

\subsection{Saddle--point analysis}
The sum over partitions that defines $Z_{X_p}$ in (\ref{totalz})
can be rewritten as
\be \label{totalz2}Z_{X_p}=\sum_R W_R^2 \,q^{(p-2) \kappa_R/2}\, 
\re^{-\ell(R)t }. \ee
where for simplicity we have absorbed the sign $(-1)^{\ell(R)p}$ in the K\"ahler parameter $t$.
To analyze this in the saddle--point limit, we will adopt a strategy used by Kostov, Staudacher and Wynter in \cite{ksw} to
analyze the partition function of Hurwitz theory (\ref{zksw}). We will introduce a ``fake" $N$ dependence in the theory, which allows
for standard large $N$ analysis, and we will then extract the genus zero result.

To see this, we first notice that (\ref{totalz2}) admits an
\textit{evocative} representation in terms of a q-deformed group
theoretical quantity of $U(N)$. Let $\{l_i\}$ be the lengths of
rows in a Young tableau, and let
$ h_i=l_i+ N-i. $
Consider the q-deformed quantity
\be _q\Omega_R=\prod_{i=1}^N {[h_i]!\over [N-i]!}, \ee
and the familiar quantum dimension of an irreducible representation
$R$ \be {\rm dim}_q R=\prod_{1\le i<j\le N} {[l_i -l_j+j-i]\over
[j-i]}. \ee The brackets denote as usual q--numbers as in
(\ref{qnumb}). Using the  equality
\be {{\rm dim}_q\, R\over _q\Omega_R}%=\prod_{\tableau{1} \in R}
%{1\over [h(\tableau{1})]}
=q^{-\kappa_R/4} W_R, \ee
we can write
\be \label{KSW1} Z_{ X_p}=\sum_R \biggl( {{\rm dim}_q\, R\over
_q\Omega_R}\biggr)^2 q^{(p-1) \kappa_R/2} \re^{-t\ell(R)}. \ee
However, the actual equality in (\ref{KSW1}) holds only when the
r.h.s  is expanded as an asymptotic series in $1/N$: this
immediately suggests to exploit the general large $N$ techniques
in investigating (\ref{KSW1}).

The planar theory can be in fact immediately analyzed along the
usual strategy originally introduced by Douglas and Kazakov for
QCD$_2$ \cite{dk}. We introduce the auxiliary `t Hooft parameter $T
=N g_s$ as in (\ref{hooft}) and continuous variables in the standard
way: \be {h_i\over N}= {l_i\over N} -{i\over N} +1 \rightarrow
\ell(x) -x+1=h(x), \ee The delicate point is to evaluate the large
$N$ limit of the deformed measure. The numerator of $_q\Omega_R$
leads to
\be \log \prod_{i=1}^N \prod_{j=1}^{h_i} (q^{h_i -j\over 2} -q
^{-{h_i -j\over 2} })^2 = 2 \sum_{i=1}^N \sum_{j=1}^{h_i} \log \,
2 \sinh g_s {h_i-j\over 2}, \ee
which becomes in the large $N$ limit
\be {2 N^2 \over T} \int_0^1 dx \int_0^{h(x)} dy \log 2 \sinh T
{h(x) -y \over 2} = {2 N^2 \over T} \int_0^1 dx \biggl( {T^2 h^2
\over 4} -{\pi^2 \over 6} + {\rm Li}_2(\re^{-T h})\biggr). \ee
The denominator of $_q\Omega_R$ cancels against the denominator of
$({\rm dim}_q\, R)^2$.  The numerator of the quantum dimension
leads to a $\sinh\, (x-y)$ interaction, as explained in
\cite{capo,abms,jm} in a related context. Then  we can write the
effective action controlling the leading large $N$ contribution as
follows
\be \ba
S&=-\int_0^1 \int_0^1 dx dy \log \Bigl| 2 \sinh {T\over 2} (h(x) -h(y))\Bigr| +{2\over T} \int_0^1 dx {\rm Li}_2 (\re^{-Th}) \\
&+ \int_0^1 dx h(x) (t- (p-1)T) + {pT\over 2}  \int_0^1 dx h^2(x)
+ (p-1){T\over 3} -{\pi^2 \over 3T} -{1\over 2} t. \ea \ee
%
%As a check of the correctness of this result we can verify that in
%the limit given by (\ref{double}) and (\ref{scaling}), we recover
%the KSW effective action, by simply using the expansion of ${\rm
%Li}_2(\re^{-T h})$:
%
%\be {\rm Li}_2(\re^{-Th})={\pi^2 \over 6} + h T \bigl( \log\, T +
%%\log\, h -1\bigr) + {\cal O}(T^2). \ee
The planar theory can
be, thus, understood as coming from a matrix model: we notice, in fact, that the
effective action can be derived from a Chern--Simons--like matrix model \cite{LHlec}
with a potential $V(h)$ of the form
\be V(h) = {2\over T} {\rm Li}_2(\re^{-Th}) + (t- (p-1)T) h +
{p\,T\over 2}  h^2, \ee
and the saddle--point equation is simply
\be \label{saddle} \int dh' \rho(h') \coth {T\over 2} (h-h') = ph
+ {2 \over T} \log(1-\re^{-Th})+ {t\over T} -(p-1), \ee
where the density $\rho(h)$ is defined in terms of the inverse
function $x(h)$  as follows \beq \rho(h)=-\frac{d x(h)}{d h}. \eeq
Because of the positivity constraint \be \label{constry}h_1>h_2
>\cdots h_N\ge 0 \Rightarrow h(x)\ge 0, \ee which the
Young tableaux variables $h_i$ must satisfy, the support of 
$\rho(h)$ will be chosen in the interval $[0,a]$.

The above equation
can be related to a standard Riemann-Hilbert problem: to this aim
we first introduce $x=1-hT$, $z=1-h'T$, and then we pass to exponential
variables
 \be
 s=\re^x, \quad y=\re^z.
 \ee
In terms of these variables the saddle--point equation reads
 \be
 \label{sadeq1}
 \int^{\re}_{\re^{- \beta}} {dy \over y}
 \rho(y) {s+ y\over s-y} = p \log s - (t+1) +(p-1)(T-1) - 2\log (1-\re^{-1} s),
 \ee
with $-\beta=1-T a $. The normalization of $\rho$ is now
\be \int_{\re^{- \beta}}^{\re}dy { \rho(y) \over  y }=T. \ee
The support of the density $\rho(s)$ comes from the original
tableau variables $h$, i.e. $[0,a]$.

To solve the saddle point equation (\ref{sadeq1}) we need a further
ingredient, namely we have to choose an ansatz for the density
$\rho(s)$. In order to recover the large radius expansion in ${\rm
e}^{-t}$ the analogy with QCD$_2$ \cite{ct} suggests to choose a
chiral, one--cut ansatz: for $x\in [-\beta,-\gamma]$ ($-\gamma<1$)
the $\rho(s)$ is arbitrary, while for $x\in[-\gamma,1]$ we require
$\rho$ to be equal to $1$. With this assumption,  we easily arrive
at our final form for the saddle-point equation
 \be
 \label{RHproblem1}
 \int_{\re^{-\beta}}^{\re^{-\gamma}}{dy \over y}  {\rho(y) \over s-y}= {p \over 2 s} \log s -{t -p(T-1) \over 2s} -{1\over s} \log (1-\re^{\gamma}
 s).
\ee

Notice that in principle this equation and its solution depend on
the parameter $T=Ng_s$, therefore the genus zero free energy of this
model $F_0(T,p,t)$ depends on $T$, as well as on $p$ and $t$.
However, the parameter $N$ was introduced by hand and does not
appear in the original model. Consistency of our procedure requires
then
\be
\label{consis}
F_0(T,p,t) ={1\over T^2}F_0(T=1, p,t) ={1\over T^2} F^{X_p}_0(p,t).
\ee
The reason for this is that $N^2 F_0(T,p,t)$ should equal $g_s^{-2} F^{X_p}_0(p,t)$. We can indeed verify this in detail as follows.
If we perform the changes of variable
\beq
\label{T1}
s\mapsto z
=s\re^{T-1}\ \ \ \  \mathrm{and}\ \ \ \ \  y\mapsto v= y\re^{T-1},
\eeq
accompanied with the redefinitions
 \beq
 \label{T2}
  \hat\gamma=\gamma+1-T,\ \ \ \ \ \hat\beta=\beta +1-T\ \ \ \ \ \mathrm{and}\ \ \ \hat\rho(v)=\rho(v \re^{1-T}) ,
 \eeq
 the saddle point equation takes the form
\beq
 \int_{\re^{-\hat\beta }}^{\re^{-\hat\gamma}}{dv \over v}  { \hat\rho(v ) \over z-v}= {p \over 2 z} \log z -{t  \over 2 z} -
 {1\over z} \log (1-\re^{ \hat\gamma}z ).
\eeq
The normalization condition instead reduces to
\beq
 \int_{\re^{-\hat\beta }}^{\re^{-\hat\gamma}}{dv \over v}   \hat\rho(v )=-\gamma.
\eeq We can now verify (\ref{consis}) by computing the $T-$dependence of the derivative of $F_0$
with respect to $t$. In terms of the original Young tableaux
variables, we have
 \beq
-\frac{\partial F_0}{\partial t}=\int_{0}^a  \rho(h) h dh -{1\over 2}. \eeq Now
if we set $ h={1-\log(v e^{1-T})}/{T}$, with the help of the
redefinitions (\ref{T1}) and (\ref{T2}), the above equation can be
cast in the following form \beq \label{pipoo}
\begin{split}
\frac{\partial F_0}{\partial t}
&=-\frac{1}{T^2}\left(\frac{T^2}{2}+\frac{{\hat\gamma}^2}{2}-\int^{\re^{-
\hat\gamma}}_{\re^{-  \hat\beta}} \dd v\,\rho(v)\log(v)\right)+{1\over 2}
=\frac{1}{T^2}\frac{\partial F_0^{\mathrm{{T=1}}}}{\partial t},
\end{split}
\eeq  as needed. In the following
we shall set $T=1$, since in this way the planar free energy of the above matrix model
equals the prepotential $F^{X_p}_0(p,t)$; the $T$ dependence is eventually recovered
through the relations (\ref{T1}), (\ref{T2}) and (\ref{pipoo}).

\subsection{Solving the saddle--point equation}
\begin{figure}[htb]
\begin{center}
\epsfxsize=4.0 in \epsfbox{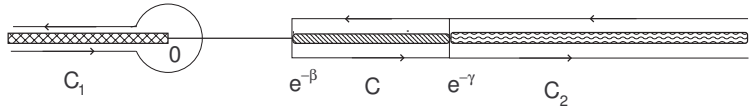}
\end{center}
\caption{The contour $C_1$ surrounds the branch cut $(-\infty,0]$ of
  $\log(z)$, $C_2$ encircles the cut $[\e^{-\gamma},\infty)$
  of $\log\big(1-e^\gamma v\big)$, while $C$
  encloses the physical branch cut $[\e^{-\beta},\e^{-\gamma}]$ %of
 %$\sqrt{(z-\e^{-\beta})(z-\e^{-\gamma})}$
 .}
\label{fg11}\end{figure}

The solution to the integral equation (\ref{RHproblem1}) can be
written in terms of the {\it effective} resolvent function \bea
\omega(z):= \int_{\e^{-\beta}}^{\e^{-\gamma}}\frac{d v}{v}~
\frac{\rho(v)}{z-v}, \eea which is given by Muskhelishvili--Migdal's
formula
 \beq
\begin{split}
\label{resfndef1} \omega(z) &=\frac{1}{2\pi i}
\sqrt{(z-\re^{-\beta})(z-\re^{-\gamma})}\oint_C \frac{dv}{(z-v)v}
\frac{\frac{p}{2}\log v
  -\frac{ t}{2}
  -\log\left(1 - v e^{ \gamma}  \right)}{\sqrt{(v-\re^{-\beta})(v-\re^{-\gamma}})}
\end{split}
\eeq where the closed contour $C$ encircles the support
$[\e^{-\beta},\e^{-\gamma}]$ of the distribution $\rho(v)$ with
counterclockwise orientation in the complex $z$-plane.
 If we choose
the square root and logarithmic\footnote{We have always defined the
logarithm function as having a branch cut along the negative axis.
This choice  implies that $\log(1-e^\gamma w)$ has a cut for $w\ge
e^{-\gamma}$.} branch cuts in (\ref{resfndef1}) as indicated in
Fig.~\ref{fg11},
 then since the integrand decays as
$v^{-3}$ at $|v|\to\infty$, we can deform the contour of
integration $C$ so that it encircles the cuts of the two
logarithms. This deformation picks up an additional contribution
from the pole at $w=z$ and we find \bea \label{resolvent1}
\omega(z)
%&\!\!\!=\!\!\!&-\frac{ 1}
%{2\pi\ii }\,\sqrt{(z-\e^{\-\gamma})(z-\e^{-\beta})}\,\left[
%\oint_{C_1}\,\frac{\dd w}{w(z-w)}~\frac{\frac{p}{2 }
%\log(w)-\frac{t}{2}-\log\left(1-{e^{\gamma}
%w}\right)}{\sqrt{(w-\e^{-\gamma})(w-\e^{-\beta})}}
%\right.\nonumber\\ &&+\left.
%\oint_{C_2}\,\frac{\dd w}{w(z-w)}~\frac{\frac{p}{2 } \log(w)-\frac{t}{2}-\log\left(1-{e^{\gamma} w}\right)}{\sqrt{(w-\e^{-\gamma})(w-\e^{-\beta})}}
%\right]+\frac{ p}{2}\,\frac{\log(z)}{z}-\frac{t}{2
%z}-\frac{\log\left(1-{e^{\gamma} z}\right)}{z}
%\nonumber\\
 &\!\!\!=\!\!\!&\left[
\frac{p}{2
z}\log\!\!\left(\!\!\frac{\left(e^{-\frac{\beta}{2}}\sqrt{z-e^{-\gamma}}
+e^{-\frac{\gamma}{2}}\sqrt{z-e^{-\beta}}\right)^2}{\left(\sqrt{z-e^{-\gamma}}
+\sqrt{z-e^{-\beta}}\right)^2}\!\right)+\frac{1}{z} \log\left(\frac{\left(1+
\frac{\sqrt{z-e^{-\beta}}}{\sqrt{z-e^{-\gamma}}}
%(e^{d^\prime}-e^{-\gamma})
\right)^2}{%(s-e^{d^\prime})
(e^{-\gamma}-e^{-\beta})}\right)\right.\nonumber\\
&&\left.-\frac{e^{\frac{\beta+\gamma}{2} } }{
z}\sqrt{(z-\e^{-\gamma})(z-\e^{-\beta})}
    %\left(  %\beta + \gamma  +
    \, \left( p\log \left(\frac{e^{\frac{\beta}{2}} + e^{\frac{\gamma}{2}}}{2}\right) -\log\left(\frac{e^{-\frac{\gamma}{2}}+
e^{-\frac{\beta}{2}}}{e^{-\frac{\gamma}{2}}-e^{-\frac{\beta}{2}}}\right) \right)
    \right]\nonumber\\
%\frac{}{} &&\left.
%+\frac{e^{\frac{\beta+\gamma}{2}}}{z}\sqrt{(z-\e^{-\gamma})(z-\e^{-\beta})}
%\log\left(\frac{e^{-\frac{\gamma}{2}}+
%e^{-\frac{\beta}{2}}}{e^{-\frac{\gamma}{2}}-e^{-\frac{\beta}{2}}}\right)\right]\nonumber\\
[5pt]&&-\frac{\gamma}{z}-\frac{t}{2 z} -\frac{t}{2 z}
\e^{(\beta+\gamma)/2}\sqrt{(z-\e^{-\gamma})(z-\e^{-\beta})}. \eea
Because of the chiral ansatz,  the boundary condition for large $z$ explicitly depends  on the extremes of the interval
and it imposes that $\omega(z)\sim -\gamma/z$.
%\beq
%\omega(z):= \int_{\e^{-\beta}}^{\e^{-\gamma}}\dd u~
%\frac{\varrho(u)}{z-u} \sim \frac{1}{z}
%\int_{\e^{-\beta}}^{\e^{-\gamma}}\dd u~\varrho(u)=-\frac{\gamma}{z}.
%\eeq
Then the  vanishing of the constant term in (\ref{resolvent1})  at infinity implies that
%\beq \frac{p }{2 }
%    \left(  \beta + \gamma  + 2\,\log \left(\frac{e^{-\frac{\beta}{2}} + e^{-\frac{\gamma}{2}}}{2}\right)  \right)-
%\log\left(\frac{e^{-\frac{\gamma}{2}}+
%e^{-\frac{\beta}{2}}}{e^{-\frac{\gamma}{2}}-e^{-\frac{\beta}{2}}}\right)+\frac{t}{2}=0
%\eeq
% which can be rewritten as
 \beq
 \label{saddle1}
 (1-p)\log \left(\frac{e^{\frac{\beta}{2}}
+ e^{\frac{\gamma}{2}}}{2}\right)- \log
\left(\frac{e^{\frac{\beta}{2}}
-e^{\frac{\gamma}{2}}}{2}\right)=\frac{t}{2}. \eeq The matching of
the subleading term in the expansion ($z^{-1}$)  instead imposes
the following endpoint equation
%\beq
%{p}\log\!\!\left(\!\!\frac{e^{-\frac{\beta}{2}}
%+e^{-\frac{\gamma}{2}}}{2}\!\right)-\log\left(\frac{
%e^{-\gamma}-e^{-\beta}}{4}\right)-{\gamma}-\frac{t}{2}=-{\gamma}
%\eeq or equivalently
\beq \label{saddle2}
(p-1)\log\!\!\left(\!\!\frac{e^{-\frac{\beta}{2}}
+e^{-\frac{\gamma}{2}}}{2}\!\right)-\log\!\!\left(\!\!\frac{e^{-\frac{\gamma}{2}}-e^{-\frac{\beta}{2}}}{2}
\!\right)=\frac{t}{2}. \eeq Our goal, in the following,  is to
solve these two equations  in a closed form. The first step is to
show that (\ref{saddle1}) and (\ref{saddle2}) can be reduced to a
polynomial equation for an auxiliary unknown $w$. First, we
introduce two intermediate variables $x$ and $y$ defined as \beq
x=\frac{e^{-\frac{\beta}{2}}+e^{-\frac{\gamma}{2}}}{2}\ \ \ \ \ \
y=\frac{e^{-\frac{\gamma}{2}}-e^{-\frac{\beta}{2}}}{2}, \eeq
which, because of their definition, have to obey the following
inequalities: $ x>0,\   y\ge 0,\  x-y=e^{-\frac{\beta}{2}}>0 $.
The saddle point equations (\ref{saddle1}) and (\ref{saddle2}) now
read \beq \frac{e^{-t/2} x^{1-p} \left(x^2-y^2\right)^p}{y}-1=0\ \
\ \ \ \ \ \ \ \frac{e^{-t/2} x^{p-1}}{y}-1=0. \eeq The second
equation can be easily solved with respect to $y$ ($y=e^{-t/2}
x^{p-1}$) and, after a trivial algebraic manipulation,  we are
left
with % one equ for $x$,
%\beq x^2 \left(1-e^{-t} x^{2
%p-4}\right)^p-1=0, \eeq
%which can be equivalently written as
\beq
\label{piro} x^{-2\frac{(p-1)^2}{p}}
\left(x^{2/p}-1\right)=e^{-t}. \eeq
The form of (\ref{piro})
suggests that the natural variable for our problem is
\be
\label{wvar}
w=x^{-\frac{2}{p}},
\ee
in terms of which (\ref{piro}) becomes a
polynomial equation
\beq
\label{ciro}
\re^{-t}=w^{(p-1)^2-1}-w^{(p-1)^2}\equiv f(w)~~.
\eeq
Because of the inequalities constraining $x$ and $y$, we are interested in the
solutions of (\ref{ciro}) that satisfy $ w>e^{-\frac{t}{p(p-2)}}$.
Once we have solved (\ref{ciro}), the endpoints are then recovered through the relation
\beq \label{endp1} \gamma = -2 \log
\left(w^{-p/2}+e^{-t/2} w^{-\frac{1}{2} (p-1) p}\right)\ \
\mathrm{and}\ \ \beta = -2 \log \left(w^{-p/2}-e^{-t/2}
   w^{-\frac{1}{2} (p-1) p}\right).
   \eeq

\medskip
\noindent The existence and the properties of such solutions can
be investigated by studying the behavior of the r.h.s. of
(\ref{ciro}) and, in particular, it is sufficient  to focus on the
region $0\le w\le 1$, since $e^{-t}$ belongs to the interval
$(0,1]$ when $t\ge 0$. In this region, for $p>2$, the derivative
of the r.h.s. of (\ref{ciro}), {\it i.e.} \beq
f^\prime(w)=w^{(p-1)^2-2}((p-2)p-(p-1)^2 w), \eeq has a unique
zero given by
\beq
\label{wc}
w_c=\frac{p(p-2)}{(p-1)^2}<1.
\eeq
Moreover it
is positive for $0<w<w_c$ and negative for $w>w_c$.
%The
%properties, for $p>2$,  of the r.h.s. of (\ref{ciro}) are
%summarized in fig. \ref{fg5}

\begin{figure}[htb]
\begin{center}
\epsfxsize=5.0 in \epsfxsize=3.5 in \epsfbox{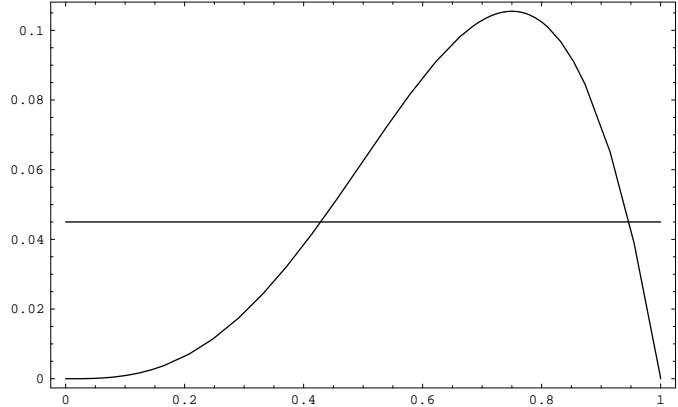}
\end{center}
\caption{\label{fg5} Plot of the r.h.s. of (\ref{ciro}) in the
region $0<w<1$. The straight line represents the constant value
$e^{-t}$.}  \end{figure}

%\begin{figure}[htb]
%\begin{center}
%\epsfxsize=5.0 in \epsfxsize=3.5 in \epsfbox{ciro.eps}
%\end{center}
%\caption{\label{fg5} Plot of the r.h.s. of (\ref{ciro}) in the
%region $0<w<1$. The straight line represents the constant value
%$e^{-t}$.}  \end{figure}
%
%\noindent
Then solutions of (\ref{ciro}) exist if and only if $e^{-t}$ is
less than the maximum located at $w=w_c$. In other words, we must
have $\re^{-t}\le f(w_c)$. This inequality is saturated by a minimum value of $t$ given by
 \beq \label{TC}  t_c=\log\left(
(p(p-2))^{p(2-p)} (p-1)^{2 (p-1)^2}\right). \eeq Above this
critical value, there are two solutions: the former is smoothly
connected to $w=1$ while the latter is smoothly connected to
$w=0$, as fig. \ref{fg5} shows.

\noindent Because of the bound $ w> e^{-\frac{t}{p(p-2)}}$, the solution which
appears  to describe the regime for large $t$ is the one near $w=1$. This solution
can be constructed as a series in $e^{-t}$:
$$\displaystyle{
w=1+\sum_{n=1}^\infty x_n \re^{-nt}.}
$$
The coefficients $x_n$ are determined,  in a closed form, by means of the Lagrange inversion
formula applied to (\ref{ciro}) (see appendix A). We obtain, after reinserting the sign factor through
$\re^{-t} \rightarrow (-1)^p \re^{-t}$,
\be
\label{serie}
w=
1+\sum_{n=1}^{\infty} \prod_{k=1}^{n-1} \Bigl( k+1-n(p-1)^2\Bigr) {(-1)^{n(p+1)} \over n!} \re^{-n t}.
\ee
An intriguing feature of this expansion, even if it is not evident, is that its coefficients
are {\it integer numbers} for integer $p$. This suggests that the relation between $w$ and $\re^{-t}$ is a 
mirror map, which is indeed characterized (for yet unknwon reasons) by integer coefficients.

\noindent
It is worth noticing that the above series can be also summed. Our equation
belongs, in fact, to a special class of algebraic equations known as
\textit{trinomial equations}. These are equations of the
form \beq w^n-a w^s+b=0,\ \ \ \ \  \textrm{with}\ n>s, \eeq
and it was shown by Birkeland (see Appendix A in \cite{birke}) that they can be
solved in terms of generalized hypergeometric functions. In our
case, we find
\beq
\label{hyperone}
w= \frac{1}{n}+\frac{n-1}{n}\
_{~n-1}\!F_{n-2}\left(\mbox{\begin{tabular}{cccc|c}
                                    $\frac{n-2}{n}$ & $\cdots$ & $\frac{1}{n}$ &$-\frac{1}{n}$ &$\zeta$\\
                                    $\frac{n-2}{n-1}$ & $\cdots$ & $\frac{1}{n-1}$ &  & \\
                                  \end{tabular}}\right),
\eeq with $n=(p-1)^2$ and $\zeta=e^{-(t-t_c)}$.

\noindent It is instructive to investigate the radius of
convergence of the  series (\ref{serie}). The simplest way of
finding it is to look directly at (\ref{hyperone}). From the
theory of hypergeometric series, we know that our series converges
only if $\zeta\le 1$, namely $t\ge t_c$: it starts diverging for
the same value $t_c$ for which the solutions to the endpoints equations stop existing. The
same result can be reached by an estimate of the asymptotic
behavior of the $x_n$ through the Stirling formula. As we will see in the next section, this is also
the radius of convergence of the prepotential.

In this analysis, we have neglected the cases $p=1,2$, which
deserve special attention. In these two cases, solutions exist for
any value of  the parameter $t$ greater than zero and the solution
can be given in terms of elementary functions: (\ref{ciro}) is, in
fact, of the first order  in $w$ both for $p=1$ and $p=2$.

At this point we should remark that the above solution can be
potentially shadowed by the breaking of the chiral ansatz. In fact,
the endpoint $\gamma$ may reach $-1$ ($h=0$) for a value $t_C \ge
t_c$. However, there are various reasons why this transition is not
relevant for our analysis. First of all, this transition depends on
the value of $T$: it exists only for small value of $T$ and  when
$T\gtrsim 1.68$, the transition disappears completely. This
indicates that the transition is not intrinsic to our original model
(\ref{totalz2}), but rather an artifact of the matrix model
description. Second, this transition is related to the
Gross--Witten--Wadia transition \cite{gw,w}, therefore it does not
characterize the expansion (\ref{KSW1}) in itself. The transition
appears only when this expansion is regarded as the strong coupling
expansion of a unitary matrix model.

\subsection{The planar free energy}
When the $(\beta,\gamma)$ are chosen so that equations
(\ref{saddle1}) and (\ref{saddle2}) are satisfied, the form of the
resolvent greatly simplifies. In fact, only the logarithmic parts
survive and we are left with \beq
\begin{split}
&\omega(z)=-\frac{\gamma}{z}-\frac{t}{2
z}+\\
&+\left[ \frac{p}{2
z}\log\!\!\left(\!\!\frac{\left(e^{-\frac{\beta}{2}}\sqrt{z-e^{-\gamma}}
+e^{-\frac{\gamma}{2}}\sqrt{z-e^{-\beta}}\right)^2}{\left(\sqrt{z-e^{-\gamma}}
+\sqrt{z-e^{-\beta}}\right)^2}\!\right)+\frac{1}{z}
\log\left(\frac{\left(1+
\frac{\sqrt{z-e^{-\beta}}}{\sqrt{z-e^{-\gamma}}}
%(e^{d^\prime}-e^{-\gamma})
\right)^2}{%(s-e^{d^\prime})
(e^{-\gamma}-e^{-\beta})}\right) \right].
\end{split}
\eeq Its jump around the cut $(\re^{-\beta},\re^{-\gamma})$
determines our genus zero density function $\rho$. We get \beq
\label{rhoq}
\rho(s)=\frac{p}{\pi}\arctan\left(\sqrt{\frac{e^{-\gamma}-s}{s-e^{-\beta}}}\right)+
\frac{2}{\pi}\arctan\left(\sqrt{\frac{s-e^{-\beta}}{e^{-\gamma}-s}}\right)-
\frac{p}{\pi}\arctan\left(e^{\frac{\gamma}{2}-\frac{\beta}{2}}\sqrt{\frac{e^{-\gamma}-s}{s-e^{-\beta}}}\right).
\eeq We now possess all the necessary ingredients for computing
the partition function at genus zero. We proceed as usual and we
first evaluate its derivative with respect to $t$. It is given by
\beq \label{part1}
\begin{split}
&\mathcal{I}=\frac{\partial({ F}_0(  t, p))}{\partial
t}=\int_{e^{-\beta}}^{e^{-\gamma}}\frac{\dd s}{s} ~\log s\, \rho(s)-
\frac{\gamma^2}{2}=-\frac{1}{2}\int_{e^{-\beta}}^{e^{-\gamma}} \dd
s~(\log s)^2\, (\rho(s))^\prime,
\end{split}
\eeq where the contribution $\displaystyle{-{\gamma}^{2}/2}$ comes
from the region where the original density is constant. The last
integral in (\ref{part1}) can be computed  in a closed form and
only in terms of  $w$.  To this purpose, we shall first change the
integration variable from $s$ to $\phi$ \beq
s=\frac{1}{2}(e^{-\gamma}+e^{-\beta}-(e^{-\gamma}-e^{-\beta})\cos\phi).
\eeq The integral then reduces to \beq
\begin{split}
\mathcal{I}%&=\int_0^\pi\dd\phi\ \mbox{$\frac{\left(e^{\beta } +e^{\gamma } -\left(e^{\beta }-e^{\gamma }\right) \cos \phi \right) (p-2)-2
%e^{\frac{\beta +\gamma }{2}} p}{4 \pi  \left(\left(-e^{\beta }+e^{\gamma }\right)
%   \cos (\phi )+e^{\beta }+e^{\gamma }\right)}
%\log ^2\left(\frac{e^{-\beta }+e^{-\gamma }-\left(e^{-\gamma }-e^{-\beta }\right) \cos \phi }{2} \right)
% $}=\\
 &=\int_0^\pi \dd\phi\  \mbox{$\left[\frac{p-2}{4\pi}-
 \frac{
e^{-\frac{\beta +\gamma }{2}} p}{2 \pi  \left(\left(e^{-\beta }-e^{-\gamma }\right)
   \cos \phi +e^{-\beta }+e^{-\gamma }\right)}\right]
\log ^2\left(\frac{e^{-\beta }+e^{-\gamma }-\left(e^{-\gamma }-e^{-\beta }\right) \cos \phi }{2} \right)
 $}
\end{split}
\eeq and it  can be exactly performed as shown in appendix B and the
result in terms of  the two endpoints is \bea \mathcal{I}&=&
%(p-2)\left( \log^2\frac{e^{-\beta /2}+e^{-\gamma /2}}{2}+\frac{1}{2}
%\mathrm{Li}_2\left(\left(\frac{e^{\beta /2}-e^{\gamma /2}}{e^{\beta /2}+e^{\gamma /2}}\right)^2\right)\right)+\nonumber\\
%&&-
%\frac{p}{2}
%\left(2\log ^2\left(\frac{\left(e^{\beta /2}+e^{\gamma /2}\right)}{2}\right)+\text{Li}_2
%\left(
%\left(
%\frac{e^{\beta /2}-e^{\gamma /2}}{e^{\beta /2}+e^{\gamma /2}}
%\right)^2
%\right)
%\right)=\\
%&=&
-\text{Li}_2
\left(
\left(
\frac{e^{\beta /2}-e^{\gamma /2}}{e^{\beta /2}+e^{\gamma /2}}
\right)^2
\right)+(p-2)\log^2\left(\frac{e^{-\beta /2}+e^{-\gamma /2}}{2}\right)-p\log ^2\left(\frac{e^{\beta /2}+e^{\gamma /2}}{2}\right).
\nonumber
\eea
By employing the expression of $\beta$ and $\gamma$ in terms
of $w$ (\ref{endp1}), we obtain
%\beq \mathcal{I}=(p-2) \log ^2(x)-p \log
%^2\left(x-e^{-t} x^{2 p-3}\right)+\text{Li}_2\left(e^{-t} x^{2
%p-4}\right). \eeq We can perform a further simplification if we
%use the equation of saddle point of $x$ and then expressing
%everything in terms of $w$. We obtain
\beq
\label{ff}
\mathcal{I}=\frac{1}{2}p(p-2) \log^2 (w)-\text{Li}_2(1-w).
\eeq
It is interesting to notice that taking another derivative of the partition function with
respect to $t$ produces an even simpler
result. Since (\ref{ciro}) fixes the derivative with respect to
$t$ of $w$ to be
\beq
\label{ciro1}
 w^\prime(t)=-\frac{(w-1) w}{(p-1)^2 w-(p-2) p},
\eeq an easy computation shows that
\beq\label{prepo} \frac{d^2
F_0}{d t^2}=-\log(w). \eeq

To obtain the actual free energy of genus zero, we have to
integrate (\ref{ff}) over $t$. This integration, however, can be
transformed in an integration over $w$ with help of (\ref{ciro1}).
We can write
%identity
%\beq \label{porro} \mathcal{F}_w(w(t),p)=\frac{1}{
%w^\prime(t)}\frac{\partial \mathcal{F}(t,p)}{\partial t }, \ \ \ \
%\mathrm{where}\ \ \ \  w^\prime(t)=-\frac{(w-1) w}{(p-1)^2 w-(p-2) p}, \ \ \  \mathrm{as~
%implied~ by~ (\ref{ciro})}.\eeq Substituting into (\ref{porro}), we get
\beq
\begin{split}
\frac{d {F}_{0}}{d w}(w,p)=-\left[\frac{p(p-2)}{w} +\frac{1}{w-1}\right]\left(\frac{1}{2}p(p-2) \log^2 (w)-\text{Li}_2(1-w)\right),
\end{split}
\eeq
which, integrated  over $w$, gives
\beq
\begin{split}
{F_0}(w(t),p)&=
-\frac{p^2(p-2)^2}{6}\log^3(w)+\text{Li}_3(1-w)+p(p-2)\log(w)\text{Li}_2(1-w)+\\
&+\frac{(p-2) p}{2} \left(\log (1-{w}) \log ^2({w})+2 \text{Li}_2({w}) \log ({w})-2 \text{Li}_3({w})+2 \zeta (3)\right),
\end{split}
\eeq
up to $p-$dependent constant.
With the help of the  polylogarithmic identities that connect polylogarithms of different arguments (see appendix B),
we can reduce  the partition function to its final form
 \beq
\begin{split}
\!\!\!\!
{F_0}(w(t),p)=&
p(p-2)  \text{Li}_3\left(1-\frac{1}{w}\right)+ (p-1)^2 \text{Li}_3(1-w)
-\frac{p}{6}  (p-2)  (p-1)^2  \log ^3(w).
\end{split}
\eeq We can also provide a closed
expansion for the prepotential ${F_0}$ as a series in $\re^{-t}$.
This is better done by working out the expansion of $\log \, w$ through Lagrange inversion
and integrating (\ref{prepo}) twice. In this way we obtain
\be
\label{foexplicit}
F_0^{X_p}(t)=\sum_{d=1}^\infty \frac{1}{  d!} \frac{1}{d^2}
\frac{((p-1)^2d-1)!}{(((p-1)^2-1)d)!} (-1)^{dp} \re^{-d t }. \ee
It is now easy to check the consistency of this result with
the direct expansion of free-energy in (\ref{foex}), 
justifying therefore {\it a posteriori} the choice of the chiral
ansatz. We notice that the partition function for $p=1,2$ greatly
simplifies and we obtain, in both cases, the partition function of
the resolved conifold, as expected, and up to a global sign due to our choice of $g_s$,
 \beq
\label{prepot12}
\begin{split}
F_0^{X_1} (t)=
-\text{Li}_3\left(\re^{-t}\right), \ \ \ \ \ \ \ F_0^{X_2}(t)=\text{Li}_3(\re^{-t}).
\end{split}
\eeq
Finally, we note that the exact expression (\ref{foexplicit}) only depends on $p$ through $(p-1)^2$, in accord with the
symmetry (\ref{psym}).

\subsection{Comparison with Hurwitz theory}

As we mentioned in section 3, the model (\ref{totalz2}) can be regarded as
a q--deformation of Hurwitz theory, which should be recovered in the limit (\ref{scaling}). This limit
can be taken order by order in the genus
expansion, as one can easily see by writing it in terms of the 't Hooft parameter $T$. We have to take
\be
 \label{double2} T
\rightarrow 0, \quad t\rightarrow +\infty, \quad p \rightarrow
\infty, \ee
in such a way that
\be \label{scaling2} pT =\tau_2, \quad (-1)^p \re^{-t} t=
T^2 \re^{-\tau_1},
 \ee
are kept fixed. If we write the free energy of Hurwitz theory (\ref{freehurwitz}) as
\be
\label{genushurwitz}
F_{\rm Hurwitz} = \sum_{g=0}^{\infty} \biggl( {N \over \tau_2} \biggr)^{2-2g} F_g^{\rm Hurwitz}(\mu),
\ee
where
\be
\mu=\tau_2^2 \re^{-\tau_1},
\ee
then in the limit (\ref{scaling2}) one has
\be
\label{limith}
p^{2-2g}F^{X_p}_g(t) \rightarrow F_g^{\rm Hurwitz}(\mu).
\ee

The planar limit of the Hurwitz model was analyzed in \cite{ksw} by
using matrix model techniques. It is easy to see that all of their
results can be recovered from the planar solution of the deformed
model. In the solution of \cite{ksw}, a crucial role is played by an
auxiliary variable $\chi$, which is related to the endpoints $[b,a]$
of the Young tableaux density through
\be \label{chi}\chi=\tau_1 + 2 \log {a-b\over 4}. \ee
The endpoint equations of \cite{ksw} lead to an equation relating $\chi$ to the parameter
$\mu$
\be\label{chimu} \mu=\chi\re^{-\chi}. \ee
The solution $\chi(\mu)$ to this equation is provided by Lambert's
$W$ function
\be
\label{lambert}
 \chi=-W(-\mu) = \sum_{k=1}^{\infty} {k^{k-1}\over k!} \mu^k,
\ee
which has convergence radius $\mu_c=\re^{-1}$. It follows from the results of \cite{ksw} that
the prepotential $F^{\rm Hurwitz}_0(\mu)$ of Hurwitz theory satisfies
\be
\label{hoexplicit}
\biggl(\mu{\partial \over \partial \mu}\biggr)^2 F_0^{\rm Hurwitz}(\mu)= \chi, \ee
and from here one can find the power series
expansion for the free-energy
\be \label{psexp} F_0^{\rm Hurwitz}(\mu)=\sum_{k=1}^{\infty} {k^{k-3} \over k!} \mu^k. \ee

Comparing the undeformed case with the deformed one, we realize
that the variable $\chi$ is playing here the role of the variable
$w$ introduced  in (\ref{wvar}). In fact in the
limit (\ref{double2})-(\ref{scaling2}) one has
\be
\label{chiw}
w-1 \rightarrow -{\chi \over p^2}.
\ee
Also, the equation (\ref{psexp}) is the limit of the equation (\ref{prepo}), and using this
fact or the explicit expansions (\ref{foexplicit}), (\ref{hoexplicit}) we can verify (\ref{limith}) for $g=0$.

We finish the section remarking that the connection with this
model will provide a useful tool in understanding the relation
between Hurwitz numbers and Gromov-Witten invariants, also {\it
beyond} genus zero, as presented in the Section 6.

\sectiono{Critical behavior in the planar limit}

In this section we will show, by looking at the genus zero solution for the Gromov--Witten invariants obtained in the previous
section, that there is a phase transition for $p>2$ with the critical exponent typical of 2d gravity.

\subsection{Critical properties of the planar free energy}

Since we have an exact expression for the genus zero
Gromov--Witten invariants at all degrees, we can analyze the
critical behavior by simply studying their asymptotic growth.  We
found,
\be
N_{0,k}={1\over k! k^2} {((p-1)^2 k -1)! \over (((p-1)^2-1)k)!},
\ee
up to a sign $(-1)^{pk}$. By using Stirling's formula, we obtain
\be
N_{0,k}\sim \re^{k  t_c} k^{-7/2}, \qquad k \rightarrow \infty,
\ee
where  $t_c$ is given in (\ref{TC}). Comparing with (\ref{asymprep}) we deduce that $t_c$ gives indeed the convergence radius of the expansion of the
prepotential around $t=\infty$, and we also deduce that
\be
\label{criticalbeh}
\gamma=-{1\over 2}.
\ee
The above results are valid for $p>2$. For $p=1,2$ the series is convergent for all $t>0$.

We will now analyze the behavior of the prepotential near the critical point.
%
%\be
%F\sim (\re^{-t_c}-\re^{-t})^{5\over 2}.
%\ee
%
%from The equation relating $w$ and $t$ is
%
%\be
%\re^{-t}=w^{(p-1)^2-1}-w^{(p-1)^2}= f(w)
%\ee
%
%The critical coupling $t_c$ is mapped to $w_c$, i.e. $f(w_c)=\re^{-t_c}$, which is
%characterized by
%
%\be
%f'(w_c)=0
%\ee
%
%and has the value
%
%\be
%w_c= \frac{p(p-2)}{(p-1)^2}.
%\ee
%
By using the explicit relation between $t$ and $w$ (\ref{serie}) we find
\be
\label{wwc}
w-w_c=A (\re^{-t_c}-\re^{-t})^{1/2} +\cdots
\ee
where $w_c$ is given in (\ref{wc}) and
\be
A={\sqrt {2}}{ (p(p-2))^ {1-(p-1)^2/2}
\over (p-1)^{3-(p-1)^2} } ={\sqrt 2} {w_c^{1-(p-1)^2/2} \over p-1}.
\ee
In order to extract the most singular part of the prepotential, we use (\ref{prepo}),
which leads to the expansion
\be
{d^2 F^{X_p} _0 \over dt^2} =-\log \, w_c -{1\over w_c} (w-w_c) +\cdots \sim - {A\over w_c} (\re^{-t_c}-\re^{-t})^{1\over 2}.
\ee
Integrating this equation, we find
\be
\label{planarsing}
F^{X_p} _0(t) \sim -{4 \over 15} {(p-1)^8 \over 4 w_c^3}  (w-w_c)^{5} \sim -{4 \over 15} {(p-1)^8 \over 4 w_c^3} A^{5} (\re^{-t_c}-\re^{-t})^{5 \over 2}.
\ee
which confirms the expected behavior (\ref{critprep}) and also gives us a precise value for the coefficient of the most singular
piece.  Again, the above holds only for $p>2$, and for $p=1,2$ the prepotential has a conifold--like behavior at $t=0$.

%It is easy to check that we recover the Hurwitz model as $p\rightarrow \infty$. One finds that
%
%\be
%w_c^{-{5\over 2} (p-1)^2 +2} \rightarrow \re^{5\over 2}, \quad x^{5\over 2}\rightarrow p^{-5} (\mu_c -\mu)^{5\over 2},
%\ee
%
%and including the overall factor $T^{-2}$ we find exactly $\tau_2^{-2}$ times the singular behavior of Hurwitz theory (\ref{hsingzero}).

If we describe the saddle--point by the density $\rho(h)$ given in (\ref{rhoq}), the critical behavior can be understood exactly as in the case of
matrix models (see for example \cite{dfgz}). This was pointed out for the limiting case of Hurwitz theory in \cite{ksw}, where $F_0^{\rm Hurwitz}(\mu)$
undergoes the same kind of
phase transition at $\mu_c=\re^{-1}$. For $t >t_c$, the
density $\rho(h)$ behaves near the endpoint $h=b$ as
\be \rho(h) -1 \sim (b-h)^{1\over 2}, \ee where $-\gamma=1-b $ and
$-\beta=1-a $.
Indeed, by expanding around this point we find
\be
\rho(h) -1=\sum_{k=1}^{\infty}  \alpha_k (t,p) (b-h)^{k-{1\over 2}},
\ee
where
\be
\alpha_1(t,p)=\frac{1}{\sqrt{a-b}} \Big(p-2 - p\sqrt{\frac{a}{b}}\Big)
\ee
If we now use the explicit expressions for the endpoints given in (\ref{endp1}) we find that, at the critical point,
\be
\alpha_1(t_c,p)=0, \quad \alpha_2(t_c, p)\not=0.
\ee
Therefore, criticality means, at the level of the density, that the leading branch cut singularity near $b$ is enhanced to
\be
\rho(h)-1\sim (b-h)^{3\over 2}.
\ee

The picture which emerges from this analysis is the following. In the large area phase $t>t_c$, the planar model is described by the one--cut ansatz of the
previous section, and the large radius expansion of the prepotential converges. At $t=t_c$ there is a phase transition controlled by the
critical exponent $\gamma=-1/2$,  and not by the conventional one $\gamma=0$
which is found in the quintic, in local $\IP^2$, and in other Calabi--Yau manifolds, as we reviewed in section 2.
This is probably due to the fact that $Z_{X_p}$ is an equivariant partition function, and our result
suggests that general equivariant partition functions like (\ref{equivptwo}) will
exhibit different critical behaviors for different choices of the equivariant parameters.
On the other hand, in the small area phase $t<t_c$ the one--cut ansatz is no longer a valid solution of the planar
model, since the equation (\ref{ciro}) does not have real solutions for $w$. This phase should be described by a chiral {\it two--cut} ansatz, as
argued in \cite{ksw} in the case of Hurwitz theory. It would be very interesting to obtain an explicit solution for this phase and to understand
its geometric meaning.

\subsection{Multicritical behavior}

We have seen that, for $p>2$, $Z_{X_p}$ exhibits critical behavior with $\gamma=-1/2$. In particular, if we regard this model as a
deformation of the Hurwitz model, we have found that the critical behavior found in \cite{ksw} at $p\rightarrow \infty$ persists for all $p>2$.

It is then natural to ask if one can find {\it multicritical behavior} a la Kazakov \cite{kazakov}, and obtain the critical behavior of the $(2,2 m-1)$ models, with $m \ge 3$.
The answer is yes, provided we turn on higher
Casimir operators. This does not have a clear interpretation in the context of topological string theory on toric
Calabi--Yau manifolds, but it is natural
to do if we view $Z_{X_p}$ as the partition function of 5d, $\CN=2$ Abelian gauge theory (the 5d version of \cite{lmn}).
We will however perform a precise
analysis only in the undeformed model (\ref{zksw}) and at the planar level. It is clear that the deformed model will have the same multicritical behavior, but it
is more difficult to solve in an explicit form.

Let us then turn on higher Casimir operators in the Hurwitz model and consider the partition function
\be
\label{hcasimirs}
Z_{\rm Hurwitz} (\tau_k) = \sum_R \biggl( {d_R \over |\ell(R)|!}\biggr)^2 N^{2\ell(R)}  \re^{-\tau_1 \ell(R)-\tau_2 \kappa_R/2N -\sum_{k\ge 3} {\tau_k \over k N^{k-1}} \CC_k}  ,
\ee
where
\be
\CC_k =\sum_{i=1}^N h_i^{k}.
\ee
Other definitions for the operators are possible. For example, using the operators ${\bf p}_k$ defined in \cite{optwo} we obtain the partition function
of topological string theory on $\IP^1$ with descendants of the K\"ahler class, and by using the operators defined in \cite{lmn} we obtain the
partition function of $U(1)$, four--dimensional $\CN=2$ supersymmetric gauge theory perturbed with Casimir operators.
In any case, with the above modification, the saddle point equation becomes
\be
\int_b^a dh' { \rho(h') \over h-h'} = {1\over 2} W'(h)+ \log(h-b),
\ee
where
\be
W'(h) =\tau_2( h -1)+  \tau_1 + \sum_{k\ge 3} \tau_k h^{k-1}.
\ee
The resolvent is
\be
\omega_0(h)={1\over 2} W'(h) -{1\over 2} M(h) {\sqrt { (h-a)(h-b)}} + \log(a-b)+\log \, h -2 \log \bigl[ {\sqrt{h-a}} + {\sqrt{h-b}} \bigr],
\ee
where
\be
M(h)=\oint_0 {dz \over 2\pi \ri} {W'(1/z) \over 1-hz} {1\over {\sqrt{(1-a z)(1-bz)}}}.
\ee
The density of eigenvalues is
\be
\rho(h)={1\over 2\pi} M(h) (h-b)^{1\over 2} (a-b)^{1\over 2}  {\sqrt{1-{h-b \over a-b}}} + {2\over \pi} \cos^{-1} \biggl( {h-b\over a-b}\biggr)^{1\over 2}.
\ee
We can expand it around the endpoint $h=b$, and we find as before
\be
\rho(h) -1=\sum_{k=1}^{\infty}  \alpha_k (\tau_k, p) (b-h)^{k-{1\over 2}}.
\ee
The $m$--th critical point is achieved when
\be
\rho(h)-1 \sim (b-h)^{m -{1\over 2}}, \qquad m\ge 2,
\ee
i.e. we have to fine--tune the $\tau_k$ in such a way that
\be
\alpha_k (\tau_k, p) =0, \quad 1\le k \le m-1.
\ee
When this is the case, we get a critical model with
\be
\gamma =-{1\over m}.
\ee
To guarantee this, $M(h)$ has to be a polynomial of degree at least $m-2$ in $h$, which we denote by $P_m(h)$,
\be
M(h) =P_m(h).
\ee
This is simply obtained by taking the first $m-2$ powers of $h-b$ in the series
\be
P_m(h)= \Biggl[ \biggl(1-{h-b \over a-b}\biggr)^{-{1\over 2}}F\Bigl( {1\over 2}, {1\over 2},{3\over 2}; {h-b\over a-b}\Bigr)\Biggr]_{m-2}.
\ee
We have used in the above that
\be
{\pi\over 2}-\cos^{-1}\, z = zF\Bigl( {1\over 2}, {1\over 2},{3\over 2}; z^2\Bigr) =\sum_{k=0}^{\infty} {(2k)! \over 2^{2k} (k!)^2 (2k +1)} z^{2k+1}.
\ee
It is now easy to extract the derivative $W_m'(h)$ of the $m$--th multicritical potential as
\be
W_m'(h)=4 \Bigl[ P_m(h) {h \over a-b} {\sqrt {\biggl( 1 -{a\over h}\biggr)\biggl( 1 -{b\over h}\biggr)}}\Bigr]_+,
\ee
where the subscript $+$ means as usual that we take the positive powers of $h$ in the expansion of the above
functions. We now list the first few critical polynomials. Denoting
\be
z={h-b \over a-b}
\ee
we have the simplified formula
\be
W_m'(z)=4\Bigl[ P_m(z) z {\sqrt {\biggl( 1 -{1\over z}\biggr)}}\Bigr]_+.
\ee
For example, one has
\be
\ba
W_2'(z)&=4z,\\
W_3'(z)&={8\over 3} z+{8\over 3} z^2,\\
W_4'(z)&={12\over 5} z +{8\over 5} z^2 + {32\over 15} z^3 ,\\
W_5'(z)&={16\over 7} z+ {48\over 35}z^2 +{128\over 105} z^3 + {64\over 35}z^4.
\ea
\ee

We then see that, in what concerns critical behavior, and at least at the planar level, the general partition function (\ref{hcasimirs}) is equivalent to the one matrix model,
since one can reach all the $(2, 2m-1)$ points.
As we will see in the next section, the double--scaling limit of the critical model where the first and the second Casimirs have been turned on
is the $(2,3)$ model (i.e. pure gravity). It is natural to conjecture that the double--scaling limit of (\ref{hcasimirs})  is
equivalent to the double--scaling limit of the one matrix model, and one can use it to describe the general $(2,2m-1)$ models coupled to 2d gravity. This should
also hold for the deformed model (\ref{zksw}) with higher Casimirs.

On the other hand, Okounkov and Pandharipande have shown in
\cite{op} that the asymptotics of the Hurwitz model (\ref{genhgf})
is equivalent to the asymptotics of the edge--of --the--spectrum
matrix model, which in turn is equivalent to topological gravity in
two dimensions. It is easy to see that the partition function
(\ref{hcasimirs}), when expanded in powers of the couplings
$\tau_k$, can be also expressed in terms of Hurwitz numbers. Indeed,
it is closely related to (\ref{genhgf}) by a nontrivial map relating
$\tau_k$ and $\tr \, V^k$. In this sense, we can regard the Hurwitz
model as an analog of the Gaussian matrix model with brane
insertions, while the model (\ref{hcasimirs}) is an analog of the
one matrix model. The relation between them and their equivalence in
the double--scaling limit should be interpreted as some sort of
open/closed duality for Hurwitz theory, in the spirit of
\cite{adkmv,mmss}.

\sectiono{Higher genus analysis and double--scaling limit}

\subsection{Explicit expressions at higher genus}
We have seen that the planar free energy of topological string theory on $X_p$ undergoes a phase transition at small volume, with the same critical exponent as 2d gravity.
We would like to understand now the behavior at higher genus. First we have to see if the $F^{X_p}_g(t)$ have a critical point at the same $t_c$, and if the critical exponent
depends on the genus as in (\ref{fgcrit}). If this is the case, we can define a double--scaled theory at the transition point capturing the all--genus behavior, and we can try
to extract the coefficients of the leading singularities at every genus in a unified way.

We have then to compute $F^{X_p}_g(t)$ from the partition function
(\ref{totalz}). Unfortunately, to the best of our knowledge, there
is no systematic way to compute corrections to the saddle--point
from sums over partitions. We have then to use a different approach.
We will in fact proceed in reverse: we will present an ansatz for
the structure of $F^{X_p}_g(t)$ which manifestly has the critical
behavior in (\ref{fgcrit}). Then, we will give evidence for the
ansatz. The main evidence comes from the $p\rightarrow \infty$
limit. In this case, and as we explained in section 3, our model is
equivalent to Hurwitz theory, and the ansatz holds thanks to the
results of \cite{gjv,gj}, which were the source of inspiration for
the conjectural expressions we will present.

Our ansatz gives closed expressions for $F_1^{X_p}(t)$ in terms of a simple logarithmic expression,
\be
\label{qansatzone}
F^{X_p}_1=-{1\over 24} \log (w-w_c) -{1\over 12} \log (p-1)+ {1\over 24} (p^2-2p+3) \log\, w,
\ee
and for the higher genus $F^{X_p}_g(t)$ as rational functions of the variable $w$
\be
\label{qansatz}
F^{X_p}_g={\CP_g(w,p) \over (w-w_c)^{5(g-1)} }, \quad \CP_g(w,p) =\sum_{i=1}^{5(g-1)} a_{g,i}(p) (w-1)^i.
\ee
The $a_{g,i}(p)$ have the form
\be
a_{g,i}(p) = { b_{g,i}(p) \over (p-1)^n},
\ee
where $n$ is a positive integer and $b_{g,i}(p)$ are polynomials in $p$ with rational coefficients. We have verified this ansatz by direct computation up to genus $4$.
In this ansatz the $F^{X_p}_g(t)$ are determined by $5(g-1)$ coefficients $a_{g,i} (p)$, which can be uniquely determined
from the genus $g$ Gromov--Witten invariants up to degree $5(g-1)$. These can be computed from the explicit
expression (\ref{totalz}) for $Z_{X_p}$. One then verifies that the
expression (\ref{qansatz}) obtained in this way reproduces correctly the Gromov--Witten invariants of higher degree.

For genus $2$, for example, one finds in this way:
\be
\ba
\label{gtwosol}
a_{2,5}(p)&={1 \over 2880} {p(p-2) \over (p-1)^2},\\
a_{2,4}(p)&=-{1\over 2880 } {12-14 p + 7 p^2 \over (p-1)^4},\\
a_{2,3}(p)&=-{1\over 2880 } { 36 - 106\,p + 161\,p^2 - 204\,p^3 +
171\,p^4 - 72\,p^5 + 12\,p^6\over (p-1)^8},\\
a_{2,2}(p)&=-{1\over 2880}{36 - 90\,p + 121\,p^2 - 60\,p^3 - 5\,p^4 + 12\,p^5 - 2\,p^6\over (p-1)^{10}},\\
a_{2,1}(p)&=-{1\over 240} {1\over (p-1)^{10}}.
\ea
 \ee
Once the coefficients $a_{g,i}(p)$ have been obtained,
one can deduce closed formulae for the Gromov--Witten invariants $N_{g,d}$ for all $d$ by using Lagrange inversion. We give the results for $g=1$ in Appendix A.

It is interesting to notice that the ansatz (\ref{qansatz}) is
very similar to the holomorphic ambiguity in the standard B--model topological string \cite{bcov}. There, one finds that $F_g(t)$ is given by a piece which is completely determined
in a recursive way by special geometry data and amplitudes at lower genera, plus an undetermined, holomorphic piece of the form
\be
f_g (z)={p_g(z) \over \Delta(z)^{2g-2}},
\ee
where $z$ is a natural coordinate in the complex structure moduli space, $\Delta(z)$ is the discriminant locus, and $p_g(z)$ is a polynomial in $z$ with unknown
coefficients. The exponent $2-2g$ in the discriminant is related to the critical exponent $\gamma=0$ characteristic of the $c=1$ behavior (\ref{cone}). In our model,
the $F^{X_p}_g(t)$ is given entirely by an analog of the holomorphic ambiguity, with a different singularity due to the different critical exponent. This also confirms the role
of $w$ as a natural mirror coordinate, and that the relation between $w$ and $\re^{-t}$ is a mirror map.

To further justify this ansatz, let us first recall the results of
\cite{gjv,gj} for the generating functionals of simple Hurwtiz
numbers: \be \label{hansatz} \ba
F^{\rm Hurwitz}_1(\mu)&=-{1\over 24} \Bigl( \log (1-\chi) + \chi\Bigr),\\
F^{\rm Hurwitz}_g(\mu) &={P_g(\chi) \over (1-\chi)^{5(g-1)}}, \quad P_g(\chi) =\sum_{i=1}^{3g-3} c_{g,i} \, \chi^i.
\ea
\ee
Here, $\chi$ is the variable introduced in (\ref{chi}), which is related to $\mu$ by (\ref{lambert}).
Moreover, there are explicit expressions for $c_{g,i}$ in terms of Hodge integrals. In particular, one has that
\be
\label{valuepg}
P_g(1) = {1\over (3g-3)!} \langle \sigma^{3g-3}_2\rangle_g,
\ee
which is a correlation function at genus $g$ in 2d topological gravity (see for example \cite{dfgz} for a review). This result will be useful later in this
section.

It is easy to see that, in the limit (\ref{double})--(\ref{scaling}), (\ref{qansatzone}) becomes $F^{\rm Hurwitz}_1(\mu)$,
 and that the ansatz (\ref{qansatz}) leads to expressions which are compatible with (\ref{hansatz}), provided
the polynomials $\CP(w,p)$ satisfy certain conditions. These conditions can be written as limiting conditions on the
coefficients $a_{g,i}(p)$:
 \be
 \label{limco}
\lim_{p\to \infty}  p^{8(g-1) -2i}a_{g,i}(p) =(-1)^i c_{g,i}.
 \ee
Using (\ref{chiw}) as well as the result
\be
w-w_c \rightarrow {1-\chi \over p^2}
\ee
in the limit (\ref{double})--(\ref{scaling}),
it is easy to see that, if (\ref{limco}) holds, then (\ref{limith}) is obtained. For example, by using the above expressions for genus two, one obtains in the $p\rightarrow \infty$ limit,
\be
{1\over p^2} F^{X_p}_2(t) \rightarrow {1\over (1-\chi)^5} \biggl( {12 \over 2880} \chi^3  + {2\over 2880} \chi^2\biggr),
\ee
which is the expression found in \cite{gjv,gj}.

\subsection{Double--scaling limit and Painlev\'e I}

Using the above results for higher genus, it is immediate to see that the ansatz (\ref{qansatz}) gives the critical behavior we are looking for. This
is a consequence of (\ref{wwc}). We want to analyze now the coefficients of the leading singularity. The behavior of the singular part of the planar
free energy (\ref{planarsing}) suggests
defining a scaled string coupling $z$ as
\be
\label{scaled}
z^{5/2} =g_s^{-2} {(p-1)^8 \over 4 w_c^3}  (w-w_c)^{5} = g_s^{-2}  {(p-1)^8 \over 4 w_c^3} A^{5} (\re^{-t_c}-\re^{-t})^{5 \over 2}.
\ee
We can now consider the double--scaled theory in which we take the limit
\be
t\rightarrow t_c, \quad g_s \rightarrow 0, \quad z \,\, {\rm fixed}.
\ee
In this limit, the total free energy of the model becomes the double--scaled free energy $F_{\rm ds}(z)$,
\be
F_{X_p} \rightarrow F_{\rm ds}(z).
\ee
Taking into account (\ref{planarsing}) and the expression (\ref{qansatzone}) for $F^{X_p}_1(t)$, we find that up to genus one
\be
\label{pertcrit}
F_{\rm ds}(z) =-{4\over 15} z^{5/2} -{1\over 48} \log \, z + \cdots.
\ee
This is, up to this order, the perturbative expansion of the 
free energy of 2d gravity, $F_{(2,3)}(z)$. We recall (see \cite{dfgz}) that $F_{(2,3)}(z)$ is determined as a
function of $z$ by the following equation,
\be
F''_{(2,3)}(z) =-u(z),
\ee
where $u(z)$, the specific heat, is a solution of the Painlev\'e I equation
\be
\label{pone}
u^2 -{1\over 6}u''=z
\ee
with the asymptotics at weak string coupling
\be
u(z) =z^{1\over 2} +\cdots, \quad z \rightarrow \infty.
\ee
This leads to the asymptotic expansion at large $z$,
\be
\label{painseries}
F_{(2,3)}(z) =-{4\over 15}z^{5/2}-{1\over 48} \log\, z +\sum_{g\ge 2} a_g z^{-5(g-1)/2},
\ee
with
\be
a_2={7\over 5560}, \quad a_3= {245 \over 331776}, \quad a_4={259553\over 159252480},
\ee
an so on.

In view of the above results for genus $g=0,1$, it is natural to conjecture that the double--scaled free energy of topological string
theory on $X_p$ equals the free energy of 2d gravity, at least in the genus expansion
\be
\label{conjone}
F_{\rm ds}(z) =F_{(2,3)}(z).
\ee
This means that the coefficient of the most singular part of $F^{X_p}_g(t)$ is identical to the coefficient $a_g$ in (\ref{painseries}), when expressed in terms of
the variable $z$ defined in (\ref{scaled}). In terms of the ansatz (\ref{qansatz}), in order to verify the conjecture (\ref{conjone}) we have to verify that the
polynomials $\CP_g(w,p)$ have the following value at the critical point $w=w_c$
\be
\label{conj}
\CP_g(w_c,p)=\biggl(4 {w_c^3 \over (p-1)^8}\biggr)^{g-1}  a_g, \quad g\ge 2.
\ee
It is easy to check with (\ref{gtwosol}) that this is indeed the case for $g=2$, and we have verified (\ref{conj})
up to genus four. We can indeed provide evidence for (\ref{conj}) at all genus in the limit $p\rightarrow \infty$. Taking into account the overall factors of $p$, one easily finds that
in this limit the conjectured equality (\ref{conj}) becomes
\be
P_g(1) = 4^{g-1} a_g, \quad g\ge 2,
\ee
where $P_g(\chi)$ are the polynomials that appear in the Goulden--Jackson--Vakil expressions for $F_g^{\rm Hurwitz}$ in (\ref{hansatz}). But due to (\ref{valuepg}) this
reduces to checking the following equality for 2d topological gravity correlators,
\be
{1\over (3g-3)!} \langle \sigma_2^{3g-3} \rangle_g =4^{g-1}a_g, \quad g\ge 2,
\ee
which indeed was shown to be true in \cite{iz}. This verifies our conjecture in the limiting case $p \rightarrow \infty$ at all genus.

To conclude this section, we have provided strong evidence that the
critical behavior of topological string theory on $X_p$ at small
radius is controlled by the Painlev\'e I equation for all $p>2$ (the
dependence on $p$ only enters in the normalization factor relating
$z$ to the ``bare" couplings $t, g_s$). Therefore the double--scaled
theory at the transition is 2d gravity (i.e. the $(2,3)$ model).
This is in contrast to conventional topological strings, which
undergo a phase transition at the conifold point controlled by $c=1$
string theory at the self--dual radius. The possibility that
different kinds of singularities in Calabi--Yau manifolds lead to
different universality classes was pointed out at the end of
\cite{ghv}. Here we have found a concrete realization of this
possibility.

\sectiono{Non--perturbative proposal and 2d gravity}

\subsection{The non--perturbative proposal and its critical properties}

Topological string theories, as well as ordinary string theories, are defined in principle only perturbatively, in a genus by genus
expansion, and it is natural to ask whether a nonperturbative definition is possible. In the case of the backgrounds that we are considering,
i.e. Calabi--Yau manifolds of the type $X_p$, it was proposed in \cite{aosv} based on previous works \cite{osv,v} that such a definition
can be given in terms of a q--deformed version of two--dimensional Yang--Mills theory with gauge group $U(N)$.
The strong coupling expansion of this partition function is given by the expression
\be
Z_{\rm qYM}=\sum_R \bigl( {\rm dim}_q R\bigr)^2 q^{p C_2(R)/ 2} e^{i \theta \ell(R)}.
\label{zq}
\ee
Here, $R$ is a tableau representing an irreducible representation of $U(N)$, $C_2(R)=\kappa_R + N \ell(R)$, and
the quantum dimension of $R$ is given by
\be
{\rm dim}_q R=\prod_{1\le i<j\le N} {[l_i -l_j+j-i]\over [j-i]}.
\ee
In \cite{aosv} it was shown that the asymptotic expansion of (\ref{zq}) can be written as
\be
\label{factor}
Z_{\rm qYM}=\sum_{l \in \IZ} \sum_{R_1, R_2}  Z_{R_1 R_2}^{+}(t+ p g_s l) Z_{R_1 R_2}^{-}(\bar t - p g_s l).
\ee
In this expression, $t$ is given in terms of the gauge theory data as
\be
\label{kahlerpar}
t={1\over 2} (p-2) Ng_s -\ri \theta.
\ee
The amplitude $Z_{R_1 R_2}^{\pm}$ is a perturbative {\it open} string amplitude on the $X_p$ Calabi--Yau and in the 
presence of two stacks of D--branes, labeled by the representations $R_1, R_2$. An 
explicit expression for $Z_{R_1 R_2}^{\pm}$, written in terms of the topological
vertex, can be found in \cite{aosv}. When
$R_1=R_2=\bullet$ is the trivial representation, one recovers the perturbative {\it closed} string amplitude
\be
\label{removopen}
Z_{\bullet \bullet}^{\pm}=Z_{X_p},
\ee
up to minor overall factors that do not affect the worldsheet instanton expansion. Therefore, 
the nonperturbative completion (\ref{zq}) gives the square of the perturbative
answer (as expected from \cite{osv}), plus some extra insertions of branes.

Later on, it was noticed in \cite{capo,abms,jm} that, at large $N$,
the partition function (\ref{zq}) (or more precisely, its zero
charge sector $l=0$) has a third--order phase transition for $p>2$
which is qualitatively similar to the Douglas--Kazakov transition of
2d Yang--Mills theory. In terms of the parameter $t$ which is
identified with the K\"ahler parameter of the perturbative theory,
and for $\theta=0$, the critical point is given by
\be
\label{npradius}
t_{\rm np}(p)={1\over 2} p(p-2) \log \biggl( 1 + \tan^2 \Bigl( {\pi \over p}\Bigr) \biggr).
\ee
Based on the analysis of \cite{t}, we expect that this critical value gives the radius of convergence of the strong coupling
expansion of (\ref{zq}), in the same way that (\ref{TC}) gives the radius of convergence of the perturbative theory.

\begin{figure}[!ht]
\leavevmode
\begin{center}
\epsfysize=5cm
\epsfbox{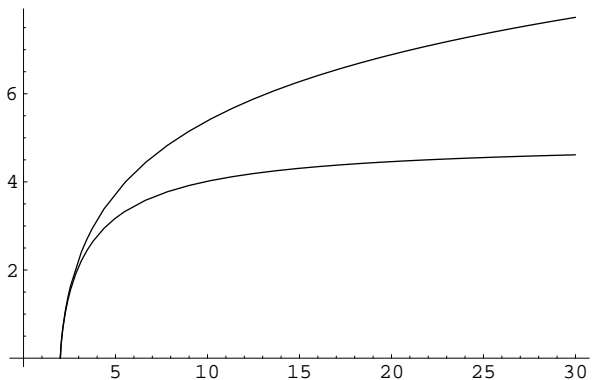}
\end{center}
\caption{The curve in the top shows the critical value of the K\"ahler parameter in the perturbative theory, which goes as $2\log \, p$ for
large $p$. The curve in the bottom shows the critical value in the q--deformed 2d Yang--Mills theory, which asymptotes $\pi^2/2$ as $p$ goes to
infinity.}
\label{radii}
\end{figure}

Therefore, {\it both} the perturbative theory, defined by $Z_{X_p}$, and the nonpertubative definition proposed in \cite{aosv}, undergo a phase transition
at small radius for $p>2$. It is interesting to compare the critical behaviors as a further probe of the proposal of \cite{osv,aosv}. The first things to compare are the
radii of convergence $t_c(p)$ and $t_{\rm np}(p)$, as a function of $p$. In \figref{radii} we plot the curves (\ref{TC}) and (\ref{npradius}) for $2<p<30$. In the perturbative theory, one has
\be
t_c \rightarrow 2 \log\, p, \quad p \rightarrow \infty,
\ee
while
\be
t_{\rm np} \rightarrow {\pi^2 \over 2},  \quad p \rightarrow \infty.
\ee
Therefore, at large $p$, the proposed nonperturbative
completion has better convergence properties as a function of $t$ at small radius.

Another interesting difference between both models is the
universality class of the theory at the transition. We have seen in
the previous section that one can define a double--scaled theory at
the critical point, and further analysis shows that this theory is
2d gravity and it is governed by the Painlev\'e I equation. A
similar study in the case of q--deformed 2d Yang--Mills theory is
still lacking, but the {\it undeformed} theory (which appears as
$p\rightarrow \infty$, as explained in \cite{capo,abms,jm}) has been
analyzed from that point of view in \cite{gm}. It turns out that
there is a well--defined double--scaled theory as well, but this
time it is in the universality class of the Gross--Witten--Wadia
unitary matrix model \cite{gw,w}. The double--scaled free energy $F(z)$ is given by 
$F''(z)=v(z)^2/4$, where $v(z)$ satisfies the Painlev\'e II equation \cite{ps},
\be
\label{ptwo}
2v'' - v^3 +z v=0.
\ee
As argued in \cite{kms}, this also describes the universality class of pure 2d supergravity.
As in the case of the perturbative theory, where the universality class is the same for all $p>2$,
we expect that q--deformed 2d YM has a double--scaled theory described by Painlev\'e II for all $p>2$. Some evidence
for this was given in \cite{abms},
based on the behavior of the
instanton suppression factor. It was shown by Crnkovic, Douglas and Moore \cite{cdm} 
that the Painlev\'e II equation (\ref{ptwo}) has a unique real, pole--free
solution which gives a non--perturbative solution of the double--scaled Gross--Witten--Wadia unitary matrix model and 
of 2d supergravity \cite{kms}. This solution, which we will call $v_{\rm CDM}$, is also 
the correct nonperturbative description of the double--scaled theory at the critical 
point of q--deformed 2d Yang--Mills. 

We therefore reach the conclusion that perturbative topological string
theory on $X_p$ exhibits a phase transition in the universality class of 2d gravity, while
the proposed nonperturbative completion exhibits a phase transition in the universality class of 2d supergravity. 
In the remaining of this section, we will present some comments and speculations on the 
possible implications of this fact for 2d gravity theories\footnote{We would like to thank Clifford Johnson for 
pointing out to us the relevance of \cite{djm,djmw} to our work, and for an illuminating email exchange. He informed us 
that he had previously envisaged the possibility that the results of \cite{djm,djmw} might be relevant in topological 
string theory.}.

\subsection{Implications for 2d gravity}

It is well--known (see again \cite{dfgz}) that the Painlev\'e I equation does not
define 2d gravity beyond the perturbation regime, since the
resulting series for the specific heat is not Borel summable. There have been various 
proposals to provide a nonperturbative definition of 2d gravity which agrees with the 
perturbative expansion (\ref{painseries}) at large $z$. A particularly interesting proposal was made 
by Dalley, Johnson and Morris in \cite{djm}. By requiring that the KdV flows of 2d gravity hold nonperturbatively they 
found a modified string equation,
\be
\label{djmstring}
u \CR^2 -{1\over 2} \CR \CR'' +{1\over 4} (\CR')^2=0, \quad \CR=u^2 -u''/3-z.
\ee
Notice that $\CR=0$ is just the Painlev\'e I equation, with a different normalization than (\ref{pone}). In this normalization, 
the free energy computed from $u$ is actually twice the free energy of pure gravity (see 
for example \cite{dfgz}). It can be shown that the above equation has a smooth, real solution $u_{\rm DJM}$, which at $z\rightarrow \infty$ 
reproduces the genus expansion obtained from Painlev\'e I, and vanishes as $1/z^2$ when $z\rightarrow -\infty$. 
This solution was proposed in \cite{djm} as a nonperturbative definition 
of 2d gravity. 

The equation (\ref{djmstring}) can be generalized by including an extra parameter $\Gamma$, as follows:
\be
\label{djmstringopen}
u \CR^2 -{1\over 2} \CR \CR'' +{1\over 4} (\CR')^2=\Gamma^2.
\ee
It was argued in \cite{djmw} that this parameter corresponds to the introduction of 
an open string sector. This equation has solutions that are open string generalizations of 
$u_{\rm DJM}$ and which we denote by $u_{\rm DJM}^{\Gamma}$. For $z \rightarrow -\infty$ they behave as
\be
u_{\rm DJM}^{\Gamma}(z) \sim {\Gamma^2 -{1\over 4}  \over z^2}. 
\ee

It was shown in \cite{djmw} that any solution $v$ 
to the Painlev\'e II equation (\ref{ptwo}) can be mapped to 
a solution $u$ of (\ref{djmstringopen}) with 
$\Gamma=\pm1/2$, through the so--called Miura transformation (see also \cite{hmpn}). In particular, the solution 
$v_{\rm CDM}$ maps under the Miura transformation to the solutions $u_{\rm DJM}^{\Gamma=\pm 1/2}$. Moreover,  if we denote by 
$Z_{\rm PII}$ and 
$Z_{\Gamma}$ the partition functions associated to the solutions related by Miura 
transformation, we have the 
relation
\be
\label{critfactor}
Z_{\rm PII}=Z_{\Gamma={1\over 2}} Z_{\Gamma=-{1\over 2}} .
\ee
Notice that this is very similar to (\ref{factor}), and in both cases one needs to introduce open 
string sectors (represented in (\ref{factor}) by the D--branes and here by the $\Gamma$ parameter). This seems to indicate that the relation (\ref{factor}) 
between the perturbative topological string amplitude and its nonperturbative completion is inherited as the relation (\ref{critfactor}) 
between their critical counterparts. 

Recall that the proposal 
of \cite{aosv,osv,v} is to regard q--deformed Yang--Mills as a nonperturbative completion of topological string theory on $X_p$. If we focus 
on the critical points of the corresponding theories, the proposal indicates that the nonperturbative completion of the 2d gravity 
perturbative expansion (\ref{pertcrit}) should be associated to the solution $v_{\rm CDM}$ of Painlev\'e II. In other words, the  
connection to Painlev\'e II should provide the extra nonperturbative information needed in 2d gravity. As we have just reviewed, $v_{\rm CDM}$ 
is related to the solutions $u_{\rm DJM}^{\Gamma=\pm 1/2}$ with an open string sector, as encoded in (\ref{critfactor}). 
However, the perturbative theory is obtained properly speaking by ``removing" the open sector, as in (\ref{removopen}). 
We conclude that the embedding of 2d gravity in perturbative topological string theory on $X_p$, together with the proposal of \cite{aosv,osv,v}, suggest 
that the natural nonperturbative completion of 2d gravity is indeed the solution presented 
in \cite{djm} $u_{\rm DJM}^{\Gamma=0}=u_{\rm DJM}$. 

We emphasize that, although we find this line of argumentation suggestive and compelling, it should be put on 
a much firmer ground by a better understanding of the nonperturbative proposal of \cite{aosv,osv,v} and of the 
embedding of the critical theories. 

\sectiono{An open--closed duality}

In section 3 we saw that the partition function $Z_{X_p}$ is related to a configuration involving a topological D--brane in $\IC^3$ with framing $f=p-1$.
In this section, we will point out a different relation which holds at genus zero and generalizes an open--closed duality proposed in \cite{av} in the context
of M--theory.

Consider then the open string background given by a D6 brane wrapping a Lagrangian submanifold with topology $\IS^1\times \IR^2$ in $\IC^3$ and with trivial
framing $f=0$. As explained in \cite{avone,akv}, this leads to a superpotential $W$ in four dimensions  which is given by
\be
W_{f=0} =\sum_{d=1}^{\infty} {\re^{d u}\over d^2}.
\ee
where $u$ is an open string modulus. From the point of view of the open string amplitudes $F_{g,\vec k}$ which appeared in section 3, this superpotential is
the generating functional of disk amplitudes with $g=0$ and a vector $\vec k$ with one single entry $k_d=1$. As pointed out in \cite{av}, this superpotential is
related to the free energy of the resolved conifold by
\be
\label{avdual}
W_{f=0}=-{dF^{X_1}_0 (t)\over dt},
\ee
after identifying $t=-u$. This relation between superpotential and prepotential
is typical from open/closed dualities, where the presence of the brane in the open side is traded by the
presence of flux in the closed side.

Consider now the same brane configuration but with arbitrary framing $f$. The superpotential has been computed in \cite{akv} and
reads
\be
W_f=\sum_{k=1}^{\infty} W_{f,k} \re^{k u}, \quad W_{f,k}={(-1)^{k(f+1)}\over k \, k!} \prod_{j=1}^{k-1} (kf-j).
\ee
Comparing this to our prepotential $F^{X_p}_0(t)$ given in (\ref{foexplicit}), we find
\be
\label{gendual}
W_{f} =- {dF^{X_p}_0 \over dt}
\ee
with the choice of framing
\be
\label{framingp}
f=(p-1)^2.
\ee
In particular, the genus zero Gopakumar--Vafa invariants $n_{0,d}(p)$ are given by
\be
n_{0,d}(p) ={1\over d}N_{d}(f),
\ee
where $N_d (f)$ are the open BPS invariants associated to disk instantons. Notice that this relation is very different from (\ref{openclosed}). The invariants
$N_{d}(f)$ can be written as linear combinations of BPS invariants $N_{R,g=0}(f)$ involving hook tableaux of $d$ boxes. But the open string
backgrounds involved in (\ref{openclosed}) and in (\ref{gendual}) are different, since in the first case one has framing $f=p-1$ while in the second case
one has $f=(p-1)^2$.

The result (\ref{gendual}) indicates the existence of an open--closed duality between closed string theory on $X_p$ and open string theory
on $\IC^3$ in the presence of a framed D6 brane with framing (\ref{framingp}), generalizing in this way the relation (\ref{avdual}) for
the resolved conifold $p=1$. In the $p=1$ case, the open/closed duality can be derived by lifting both configurations to
M--theory on a manifold with $G_2$ holonomy and with topology $\IS^3 \times \IR^4$ \cite{av}. For general $p$, one
should be able to lift the D6 brane with framing to M--theory and
relate it there to closed string theory on $X_p$, explaining in this way the equality (\ref{gendual}). This would produce a new and
interesting class of M--theory dualities.

\sectiono{Conclusions and open problems}

The main conclusions of our analysis are the following:

\begin{itemize}

\item We have shown that, in some backgrounds, topological string theory exhibits a critical behavior
which is not in the universality class of the $c=1$ string, and leads to a double--scaled theory which is identical
to pure 2d gravity (in perturbation theory).

\item From a more methodological point of view, we have shown that double--scaling limits can be useful to
characterize critical string theories, and they allow us to
partially capture their all--genus behavior. This is a useful
strategy to address questions related to their nonperturbative
behavior, and we have argued in this paper that the double--scaled
theory reflects aspects of the nonperturbative structure of the
original theory. In particular, we have seen that the proposal for a
nonperturbative completion of topological string theory on $X_p$
leads to a double--scaled theory with a well--defined nonperturbative
description in terms of the solution $v_{\rm CDM}$ to Painlev\'e II.

\item Conversely, we suggested that this result gives further support to the 
proposal of \cite{djm} for a nonperturbative completion of 2d gravity.

 \end{itemize}

Our work also leaves many open questions, and we list some of them here:

\begin{itemize}

\item From a technical point of view, our analysis has some gaps which should be filled. For example, it would be interesting to give a first principles
derivation of the higher genus ansatz (\ref{qansatz}), as well as a constructive way of computing the coefficients $a_{g,i}(p)$. In order to do that, one should
develop general techniques to compute higher genus corrections for sums over partitions, which would find applications in many other problems
related to topological strings and instanton counting. A systematic method to treat higher genus corrections would also make possible to derive the
precise properties of the double--scaled theory, as in the case of matrix models.

\item We have not determined the nature of the theory in the phase with $t<t_c$. In principle, this corresponds to a two--cut solution of the saddle--point equations, but
finding it is technically challenging. In this respect, it would be very interesting to develop mirror symmetry techniques to analyze this model (both at the planar level and
at higher genus). Some preliminary steps in this direction have already been given in \cite{fj}, but much remains to be done. This would also
shed light on the B--model analysis of equivariant topological strings introduced in \cite{amv}.

\item We have found evidence for an open--closed duality involving a lift to M--theory, and generalizing the observations of \cite{av}. It would be
very interesting to use $G_2$ holonomy manifolds and justify the relation between the D6 brane background with framing $f=(p-1)^2$
and the closed $X_p$ background.

\item There have been proposals for nonperturbative formulations of topological string theory on other toric manifolds, like local $\IP^2$ \cite{ajs}. It would be
very interesting to analyze the critical behavior of these models and their universality class. Based on the results of this paper, we would expect a ``well--behaved" critical theory, probably related to doubly--scaled unitary matrix models.

\end{itemize}

{\bf Note added}: After this paper was submitted, the work \cite{fjtwo} appeared, which studies the B--model realization of topological string theory 
on $X_p$. They derive the following explicit expression for the mirror map, 
\be
-q{dt\over dq}={1- (-1)^{p} (p-1)^2 q \over 1-(-1)^{p} q}.
\ee
It  can be easily seen that after setting $(-1)^{p-1}q=w-1$ the above expression becomes precisely our equation (\ref{ciro}). This confirms our 
conjecture that the relation between $w$ and $t$ is indeed the mirror map of this model. By using B--model techniques, \cite{fjtwo} also derives an 
expression for $F_1$ which agrees precisely with (\ref{qansatzone}). 

 \section*{Acknowledgments}

We would like to thank Mina Aganagic, Luis \'Alvarez--Gaum\'e,
Renzo Cavalieri, Gianni Cicuta, Ian Goulden, Antonella Grassi, 
Sergei Gukov, David Jackson, Clifford Johnson, Albrecht Klemm, Davesh Maulik, Hirosi
Ooguri, Rahul Pandharipande, and Ravi Vakil, for useful
conversations, comments, and correspondence. M.M. would like to
thank the string theory group at Caltech and the MSRI at Berkeley
for their warm hospitality while this work was in progress. L.G. and S.P.
would like to thank the Theory Unit at CERN for hospitality. The
work of N.C. is supported by a fellowship of the ``Angelo della
Riccia" foundation.

\appendix
\section{Lagrange inversion}

 The Lagrange inversion method makes possible to explicitly invert a relation between two variables in power series form. A good exposition
 can be found for example in \cite{henrici}.

 Let $z=f(w)$ be a relation that implicitly defines $w$ as a function of $z$ around $w=0$. We can assume that $f(0)=0$.
 We can invert this relation and find the inverse function
 \be
 w=f^{-1}(z)
 \ee
 as follows
 \be
 \label{lagrange}
w=\sum_{n=1}^{\infty}  {1\over n} {\rm Res} \biggl( {1\over f(w)^n} \biggr)  z^n,
 \ee
where the residue is computed around $w=0$. A slight modification of this result makes possible to compute
the function
\be
g(w)=g(f^{-1}(z))
\ee
as
\be
g(w)=\sum_{n=1}^{\infty}  {1\over n} {\rm Res} \biggl( {g'(w)\over f(w)^n} \biggr)  z^n.
\ee
We assumed again that $g(0)=0$.

\noindent
Let us now apply this procedure to the equation (\ref{ciro}) relating $t$ and the variable $w$
\be
q=w^{(p-1)^2}(w^{-1} -1),
\ee
where $q=\re^{-t}$. Set $f=(p-1)^2$, $x=w-1$. The equation defining $x$ is
\be
q=-x(1+x)^{f-1},
\ee
We want to find now $x$ (or $w=1+x$) as a function of $q$. By (\ref{lagrange}), we have to compute
\be
{(-1)^n \over n} {\rm Res} \, x^{-n} {1\over (1+x)^{n(f-1)}}= {(-1)^n \over n} {\rm Res} \, \sum_{k=0}^{\infty} {(n(f-1)+k-1)! \over k! (n(f-1)-1)!} (-1)^k x^{k-n},
\ee
which equals
\be
 {(-1)^n \over n!} \prod_{k=2}^{n} (k-nf).
\ee
Therefore,
\be
\label{www}
w=
1+\sum_{n=1}^{\infty} \prod_{k=1}^{n-1} \Bigl( k+1-n(p-1)^2\Bigr) {(-1)^n \over n!} q^n,
\ee
where the $n=1$ term is taken to be 1.

Lagrange inversion can be also used to find explicit expressions for the Gromov--Witten invariants of $X_p$ from 
(\ref{qansatzone}) and (\ref{qansatz}).
Let us calculate the genus one invariants. In terms of $x=w-1$, we have
\be
F_1=-{1\over 24} \log\biggl[ 1 +(p-1)^2 x \biggr] +{1\over 24} (p^2-2p+3) \log\, (1+x).
\ee
Setting again $f=(p-1)^2$ we have to compute the residue:
\be
-{(-1)^n \over 24 n} {\rm Res}\, \biggl( {1\over x^n} {f\over 1+f x} {1\over (1+x)^{n(f-1)}} -{1\over x^n} {f+2 \over (1+x)^{n(f-1)+1}}\biggr) .
\ee
and we easily obtain
\be
N_{1,k}={1\over 24 k}\sum_{\ell=0}^{k-1} {f^{k-\ell}\over \ell!} \prod_{j=1}^\ell  (k(f-1) + j-1) -{1\over 24} {(kf-1)! \over k! (k(f-1))!}(f+2).
\ee
We can check the limiting behavior (\ref{hurlimit}),
\be
\lim_{p\rightarrow \infty} {N_{1,k} \over p^{2k} }  ={1 \over 24 k} \sum_{\ell=0}^{k-2} {k^{\ell} \over \ell! }.
\ee
which is indeed the expected answer.
\section{Useful Integrals}
In this appendix we shall summarize some integrals that have been used  to perform the large $N$ analysis in sec. 4.

\noindent
All the the integrals appearing in the computation of the resolvent $\omega(z)$ can be reduced to the
following elementary indefinite integral
\bea
&&\int \frac{\dd w}{w-s}~\frac{1}{\sqrt{(w-\e^{-\gamma})(w-\e^{-\beta})}}\\
&&\qquad~=~-\frac{1}{\sqrt{(s-\e^{-\gamma})(s-\e^{-\beta})}}\,
\log\left(\mbox{$\frac{\left(\sqrt{(w-\e^{-\beta})(s-\e^{-\gamma})}+
\sqrt{(s-\e^{-\beta})(w-\e^{-\gamma})}~\right)^2}{(s-w)\,
\sqrt{(s-\e^{-\beta})(s-\e^{-\gamma})}}$}\right). \nonumber
\eea

\noindent
The computation of the planar free energy involves instead the following families
of definite integrals. First we have to consider
\bea
\mathcal{I}_1
 \!\!\!\!\!&=&\!\!\!\!\frac{1}{2}\int_{-\pi}^\pi \dd\phi\
\log ^2\left(\frac{e^{-\beta }+e^{-\gamma }-\left(e^{-\gamma }-e^{-\beta }\right) \cos \phi }{2} \right)=\nonumber\\
\!\!\!\!\!&=&\!\!\!\!\frac{1}{2}\int_{-\pi}^\pi \dd\phi\ \left[\log\frac{\left(e^{-\beta /2}+e^{-\gamma /2}\right)^2}{4}+
\log\left(1-\frac{e^{\beta /2}-e^{\gamma /2}}{e^{\beta /2}+e^{\gamma /2}}
e^{-i\phi}\right)+\right.\\
&&\ \ +\left.\log \left(1-\frac{e^{\beta
   /2}-e^{\gamma /2} }{e^{\beta /2}+e^{\gamma /2}}e^{i\phi}\right)
\right]^2\!\!\!=\!%\\
%&
 4\pi \log^2\frac{e^{-\beta /2}+e^{-\gamma /2}}{2}+2\pi
\mathrm{Li}_2\left(\left(\frac{e^{\beta /2}-e^{\gamma /2}}{e^{\beta /2}+e^{\gamma /2}}\right)^2\right)\nonumber
\eea
This integral is actually  computed by expanding the square, then by expanding in series the logarithms and finally integrating the
series. The final series can be summed  in terms of polylogarithms and logarithms .

\noindent
With the same technique, we can also show that
\beq
\begin{split}
\mathcal{I}_2
 &=\frac{1}{2}\int_{-\pi}^\pi \dd\phi\
\log^2 \left(1-C e^{i \phi} \right)\log\left(1-C e^{-i \phi} \right)=\\
&=-\pi  \left(\log \left(\text{C}^2\right) \log ^2\left(1-\text{C}^2\right)+2 \text{Li}_2\left(1-\text{C}^2\right) \log \left(1-\text{C}^2\right)-2
   \text{Li}_3\left(1-\text{C}^2\right)+2 \zeta (3)\right)
\end{split}
\eeq
and
\beq
\begin{split}
\mathcal{I}_3
 &=\frac{1}{2}\int_{-\pi}^\pi \dd\phi\
\log ^3\left(A-B \cos \phi \right)
=\pi \log ^3\left(\frac{A}{2}+\frac{1}{2} \sqrt{A^2-B^2}\right)+\\
&+6\pi \mathrm{Li_2}\left(\left(\frac{B }{A+\sqrt{A^2-B^2}}\right)^2\right)
\log \left(\frac{A}{2}+\frac{1}{2} \sqrt{A^2-B^2}\right)+6 \mathcal{I}_2\left(\left(\frac{B }{A+\sqrt{A^2-B^2}}\right)^2\right)
\end{split}
\eeq
\noindent
Using the result for $\mathcal{I}_3$, we can obtain
\beq
\begin{split}
 \mathcal{I}_4&=\frac{1}{2}\int_{-\pi}^\pi \dd\phi\
\log ^2\left(A-B \cos \phi \right)/(A-B \cos\phi)=\frac{1}{6}
\frac{\partial}{\partial A}\int_{-\pi}^\pi \dd\phi\
\log ^3\left(A-B \cos \phi \right)=\\
&=\frac{ \pi  \left(\log ^2\left(\frac{2 \left(B^2-A \left(A+\sqrt{A^2-B^2}\right)\right)^2}{\left(A+\sqrt{A^2-B^2}\right)^3}\right)+2
   \text{Li}_2\left(\frac{B^2}{\left(A+\sqrt{A^2-B^2}\right)^2}\right)\right)}{\sqrt{(A-B) (A+B)}},
\end{split}
\eeq
which can be used to evaluate the following integral appearing in our computation of the planar free energy
\beq
\begin{split}
&\int_0^\pi \dd\phi\
 \frac{
e^{-\frac{\beta +\gamma }{2}} }{  \left(\left(e^{-\beta }-e^{-\gamma }\right)
   \cos \phi +e^{-\beta }+e^{-\gamma }\right)}
\log ^2\left(\frac{e^{-\beta }+e^{-\gamma }-\left(e^{-\gamma }-e^{-\beta }\right) \cos \phi }{2} \right)=\\
&=\pi  \left(2\log ^2\left(\frac{\left(e^{\beta /2}+e^{\gamma /2}\right)}{2}\right)+\text{Li}_2\left(\tanh ^2\left(\frac{\beta -\gamma }{4}\right)\right)\right)
\end{split}
\eeq
The results of these integrals can be also presented in an equivalent and more useful way, if we use the identities
\beq
\log (1-x) \log (x)+\text{Li}_2(1-x)+\text{Li}_2(x)-\frac{\pi ^2}{6}=0.
\eeq
and
\beq
\text{Li}_3(w)= \frac{1}{6} \log ^3(w)\!-\!\frac{1}{2} \log (1-w) \log ^2(w)\!+\!\frac{\pi ^2}{6}  \log
   (w)-\text{Li}_3(1-w)-\text{Li}_3\left(\!\!\frac{w-1}{w}\right)\!\!+\zeta (3).
\eeq

%\section{Weak phase of the KSW model}

\end{document}